\documentclass{emulateapj}
\slugcomment{draft v2}

\shorttitle{Bimodal Long-Lasting Components in Short GRBs}
\shortauthors{Kisaka, Ioka \& Sakamoto}

\begin{document}


\title{Bimodal Long-Lasting Components in Short Gamma-Ray Bursts: Promising Electromagnetic Counterparts to Neutron Star Binary Mergers}


\author{Shota Kisaka\altaffilmark{1,3}}
\email{kisaka@phys.aoyama.ac.jp}
\author{Kunihito Ioka\altaffilmark{2}}
\email{kunihito.ioka@yukawa.kyoto-u.ac.jp}
\author{Takanori Sakamoto\altaffilmark{1}}
\email{tsakamoto@phys.aoyama.ac.jp}


\altaffiltext{1}{Department of Physics and Mathematics, Aoyama Gakuin University, Sagamihara, Kanagawa, 252-5258, Japan}
\altaffiltext{2}{Center for Gravitational Physics, Yukawa Institute for Theoretical Physics, Kyoto University, Kyoto 606-8502, Japan}
\altaffiltext{3}{JSPS Research Fellow}


\begin{abstract}

Long-lasting emission of short gamma-ray bursts (GRBs)
is crucial to reveal the physical origin of the central engine
as well as to detect electromagnetic (EM) counterparts to gravitational waves (GWs) 
from neutron star binary mergers.
We investigate 65 X-ray light curves of short GRBs,
which is six times more than previous studies, by combining both {\it Swift}/BAT and XRT data.
The light curves are found to consist of two distinct components at $>5\sigma$ 
with bimodal distributions of luminosity and duration, 
i.e., extended (with timescale $\lesssim10^3$ s)
and plateau emission (with timescale $\gtrsim10^3$ s), 
which are likely the central engine activities but not afterglows. 
The extended emission has an isotropic energy comparable to the prompt emission, 
while the plateau emission has $\sim0.01-1$ times of that energy. 
A half (50\%) of our sample has both components,
while the other half is consistent with having both components.
This leads us to {\it conjecture} that almost all short GRBs have both the extended and plateau emission.
The long-lasting emission can be explained by the jets from black holes with fallback ejecta, 
and could power macronovae (or kilonovae) like GRB 130603B and GRB 160821B. 
Based on the observed properties, we quantify the detectability of EM counterparts to GWs,
including the plateau emission scattered to the off-axis angle,
with {\it CALET}/HXM, {\it INTEGRAL}/SPI-ACS, {\it Fermi}/GBM, {\it MAXI}/GSC, 
{\it Swift}/BAT, XRT, future {\it ISS-Lobster}/WFI, {\it Einstein Probe}/WXT, and {\it eROSITA}.

\end{abstract}





\section{INTRODUCTION}
\label{sec:introduction}

The physical origin of short gamma-ray bursts (GRBs) remains unknown despite intensive studies 
\citep[e.g., ][]{N07, LR07}.
The leading model for a progenitor of short GRBs is neutron star (NS) binary mergers 
including NS-NS and black hole (BH)-NS binaries \citep[e.g., ][]{NPP92}, 
as inferred from the short emission timescale, the energetics, 
and the wide variety of their host galaxy type \citep{B14}. 
However, there is no smoking-gun evidence for the merger hypothesis yet.

The best way to verify the merger hypothesis is to detect coincident gravitational waves (GWs). 
Recently, the GW astronomy has begun since the direct detections of GWs from BH-BH binaries, GW 150914, 
LVT 151012, GW 151226, and GW 170104 by Advanced LIGO \citep{Abbott+16a, Abbott+16b, Abbott+16c, Abbott+17}. 
An NS binary merger is also associated with the GW emission that is detectable by the current GW detectors 
\citep{Abbott+16d}. 
Simultaneous detection of GW and electromagnetic (EM) emission, particularly the association to short GRBs, 
will provide valuable information for understanding the progenitor of short GRBs \citep[e.g., ][]{MB12, FMH17}. 

Activities of the central engine of short GRBs continue much longer timescale ($\gg1$ s) than 
the duration of the prompt emission ($\lesssim 1$ s) \citep[e.g., ][]{Barthelmy+05, IKZ05, NB06}.
{\it Swift}/BAT has detected more than 100 short GRBs, 
and $\sim70\%$ of them have been also detected by {\it Swift}/XRT
\footnote{https://swift.gsfc.nasa.gov/archive/grb\_table/}. 
Some {\it Swift}/XRT light curves show a long duration with rapid flux decline, 
which is only produced by activities of the central engine
\citep[e.g., ][]{IKZ05}.
By investigating the properties of the long-lasting components, we can obtain a clue to 
the central engine of short GRBs, and improve the strategies for the simultaneous detection with GW emission. 

One of the long-lasting activities in short GRBs is the extended emission with timescale $\sim100$ s, 
which is much longer than typical accretion timescale \citep[e.g., ][]{Barthelmy+05, NB06}. 
In some bursts, the fluence of the extended emission is comparable to or even higher than that of the initial pulse 
\citep[e.g., ][]{Per+09}. 
The fraction of bursts with the extended emission in the $\gamma$-ray band ($>15$ keV) 
is $\sim$2-25\% \citep[e.g., ][]{NGS10, Sakamoto+11, BKG13, KBGL15, Lien+16}. 
However, as shown later, a larger number of the extended emission component may be missed 
in the {\it Swift}/BAT band (15-150 keV), 
because some bursts show extended emission only in the {\it Swift}/XRT band 
\citep[0.3-10 keV; ][]{Row+13, Kag+15, Lu+15, Lu+17}.

In addition, short GRBs with extended emission in the $\gamma$-ray band also have plateau emission 
with timescale $\sim10^3-10^4$ s in the {\it Swift}/XRT band \citep{GOWR13, GOW14}.
Some theoretical models for the plateau emission suggest an activity of the central engine such as 
a relativistic jet from a BH with a typical NS magnetic field $\sim10^{12}$ G \citep{KI15} or a pulsar wind
from a highly magnetized ($\sim10^{15}-10^{16}$ G) and rapidly rotating ($\sim1$ ms) NS \citep[e.g., ][]{GOW14, GWGO17}. 
Note that some of bursts without extended emission in the $\gamma$-ray band also show the plateau-like emission component  
\citep{Row+13, Lu+15, Lu+17}.  

In order to increase the detectability of EM counterparts to GWs from NS binary mergers,  
it is important to understand the properties of the long-lasting components.  
For example, their luminosity function is necessary to estimate the integration time of the follow-up observations.
Comparing the duration distribution with the required integration time will determine 
the maximum number of pointing observations.
Therefore, it is important to investigate the statistical properties of the long-lasting components from the current 
observational data in order to improve the observational strategies for the EM counterparts. 

The properties of the long-lasting components are also important to characterize the interaction between the jets  
and the merger ejecta surrounding the central engine \citep[e.g., ][]{MQT08, BMTQ12}. 
In particular, nearly isotropic emission is anticipated through the interaction 
\citep[e.g., ][]{MP14, Nakamura+14, KIT15, KIN15, HP15, SZG16, LDMW16, GNP17}. 
Although the plateau emission may arise from a collimated relativistic jet, 
a significant fraction of the emission could be scattered to wider solid angle by the merger ejecta 
during the plateau emission activity \citep{KIN15}.
Then, the properties of plateau emission are necessary to estimate those of the scattered component. 
In addition, the extended and plateau emission activities could heat the merger ejecta, 
which is observed as a macronova \footnote{We use the term ``macronova'' 
as a thermal radiation from the merger ejecta of NS binaries, whatever the energy source is, like a supernova.}
(or kilonova) in the optical and infrared bands \citep{K05, YZG13, KIT15, KIN15}. 
The emission could be brighter than the macronova heated by the decay of 
the heavy elements \citep[e.g., ][]{LP98, K05}, in particular $r$-process elements 
\citep[e.g., ][]{Metzger+10, KBB13, TH13}, which is widely discussed. 
The luminosity of the engine-powered macronova is determined by the properties of 
the long-lasting components \citep{KIT15, KIN15}. 

Currently, the number of short GRBs with both the extended and plateau emission is only $\sim10$,  
whose extended emission was detected in the {\it Swift}/BAT band \citep{GOW14, KI15}. 
The number is too small to statistically characterize the properties of the extended and plateau emission. 
On the other hand, some bursts without the BAT-detected extended emission actually show the features
of the extended and plateau emission, that are flat flux evolution and rapid decline in the {\it Swift}/XRT band 
instead \citep{Row+13, Lu+15, Lu+17}. 
In fact, the fraction of the bursts with extended emission tends to be higher for softer threshold energy 
\citep{NGS10, BKG13}.
Hence, both the {\it Swift}/BAT and XRT bands should be used to identify the extended and 
plateau emission components \citep{Kag+15}. 
Then, the sample size of the extended and plateau emission components becomes large. 
With a large sample, we can investigate whether the extended and plateau emission is two distinct components 
or not, and whether all short GRBs have both components following the prompt emission or not. 

In this paper, we investigate the light curves of 65 short GRBs with the sufficient {\it Swift}/XRT data 
to characterize the statistical properties of both extended and plateau emission. 
Using a phenomenological model, we extract the long-lasting components, the extended and plateau emission, 
from the observed light curves. 
In Section 2, we describe our sample and the light curve model. 
In Section 3, we provide the results of the obtained luminosity and duration of the extended and plateau emission, 
and show that the distributions are bimodal. 
In Section 4, we discuss the detectability of the long-lasting components as an EM counterpart to GW from NS binary mergers. 
In Section 5, we discuss implications for theoretical models based on the obtained properties of 
the long-lasting emission components. 
Conclusions and discussion are provided in Section 6.

\section{SAMPLE AND MODEL}
\label{sec:sample}

\begin{table*}
 \begin{center}
  \begin{tabular}{lclcc}
\multicolumn{5}{c}{TABLE 1 Short GRB Samples} \\ \hline
Name & Redshift & Reference & Extended emission & Plateau emission \\ \hline
050509B       & 0.2249 & \citet{Prochaska+05} &            & \checkmark   \\
050724$^\ast$  & 0.257 & \citet{Berger+05}     & \checkmark & \checkmark   \\
051210        & (0.72) &                      &\checkmark &              \\
051221A       & 0.5464 & \citet{Soderberg+06} & \checkmark & \checkmark   \\
051227$^\ast$  & 0.8   & \citet{D'Avanzo+09}   & \checkmark & \checkmark   \\
060313         & (0.72) &                     &\checkmark & \checkmark   \\
060614$^\ast$  & 0.1254 & \citet{DellaValle+06} & \checkmark & \checkmark   \\
060801        & 1.1304 & \citet{Berger+07}    & \checkmark &             \\
061006$^\ast$  & 0.4377 & \citet{Berger+07}    & \checkmark & \checkmark   \\
061201         & 0.111  & \citet{B06}         & \checkmark & \checkmark   \\
061210$^\ast$  & 0.4095 & \citet{Berger+07}    &\checkmark & \checkmark   \\
070714A        & 1.58 & $^{\rm a}$             &          & \checkmark   \\
070714B$^\ast$ & 0.9224 & \citet{Cenko+08}    & \checkmark & \checkmark   \\
070724A        & 0.4571 & \citet{B09}         & \checkmark & \checkmark   \\
070809         & 0.2187  & \citet{Perley+08}  &             & \checkmark   \\
071227$^\ast$   & 0.381  & \citet{D'Avanzo+09} & \checkmark & \checkmark   \\
080123$^\ast$  & 0.495 & \citet{LB10}          & \checkmark & \checkmark   \\
080426         & (0.72) &                      &         & \checkmark   \\
080503$^\ast$   & (0.72) &                     & \checkmark &             \\
080702A        & (0.72) &                     & \checkmark &              \\
080905A        & 0.1218 & \citet{Rowlinson+10} & \checkmark &              \\
080919         & (0.72) &                     & \checkmark &              \\
081024A        & (0.72) &                     & \checkmark &              \\
081226A        & (0.72) &                     &           & \checkmark   \\
090426         & 2.609  & \citet{Levesque+10} & \checkmark & \checkmark   \\
090510         & 0.903  & \citet{McBreen+10}    & \checkmark & \checkmark   \\
090515         & (0.72) &                       & \checkmark &              \\
090621B        & (0.72) &                       &          & \checkmark   \\
091109B        & (0.72) &                      & \checkmark & \checkmark   \\
100117A        & 0.915  & \citet{Fong+11}     & \checkmark &              \\
100625A        & 0.452  & \citet{Fong+13}     & \checkmark & \checkmark   \\
100702A        & (0.72) &                     & \checkmark &              \\
100724A        & 1.288  & \citet{Thoene+10} & \checkmark & \checkmark   \\
101219A        & 0.718  & \citet{Fong+13}   & \checkmark &              \\
110112A        & (0.72) &                   &            & \checkmark   \\
111020A        & (0.72) &                   &            & \checkmark   \\
111117A        & 1.31 & \citet{Sakamoto+13} & \checkmark & \checkmark   \\
111121A$^\ast$ & (0.72) &                   & \checkmark & \checkmark   \\
120305A        & (0.72) &                  & \checkmark & \checkmark   \\
120521A        & (0.72) &                  & \checkmark &              \\
120804A        & 1.3 & \citet{Berger+13}  & \checkmark & \checkmark   \\
121226A        & (0.72) &                 & \checkmark & \checkmark   \\
130603B        & 0.3564 & \citet{deUgartePostigo+14} & & \checkmark   \\
130912A        & (0.72) &                 &           & \checkmark   \\
131004A        & 0.717 & \citet{CLB13}    & \checkmark & \checkmark   \\
140129B        & (0.72) &                 &            & \checkmark   \\
140516A        & (0.72) &                 & \checkmark & \checkmark   \\
140903A        & 0.351  & \citet{Troja+16} &           & \checkmark   \\
140930B        & (0.72) &                  & \checkmark & \checkmark   \\
150120A        & 0.460  & \citet{CF15} & \checkmark &              \\
150301A        & (0.72) &                  & \checkmark &              \\
150423A        & 1.394  & \citet{Malesani+15}   & \checkmark   \\
150424A$^\ast$ & 0.30   & \citet{Castro-Tirado+15} & \checkmark & \checkmark   \\
150831A        & (0.72) &                  & \checkmark & \checkmark   \\
151127A        & (0.72) &                  &          & \checkmark   \\
151229A        & (0.72) &                  & \checkmark & \checkmark   \\
160408A        & (0.72) &                  & \checkmark &              \\
160411A        & (0.72) &                  &           & \checkmark   \\
160525B        & (0.72) &                  & \checkmark & \checkmark   \\
160601A        & (0.72) &                  &           & \checkmark   \\
160624A        & 0.483  & \citet{CL16}  & \checkmark &              \\
160821B        & 0.16   & \citet{Levan+16}  & \checkmark & \checkmark   \\
160927A        & (0.72) &                   & \checkmark & \checkmark   \\
161004A        & (0.72) &                   & \checkmark & \checkmark   \\
170127B        & (0.72) &                   & \checkmark & \checkmark   \\ \hline
\multicolumn{4}{l}{$^{\rm a}$ http://www.astro.caltech.edu/grbhosts/redshifts.html}%
 \label{tab:parameter}
  \end{tabular}
 \end{center}
\end{table*}

In this paper, we refer to bursts as short GRBs if $T_{90}\le2$ s \citep{Kou+93}, 
where $T_{90}$ corresponds to the duration that contains 90\% of the burst fluence measured 
by the {\it Swift}/BAT instrument (15-150 keV).
The short GRB data sample was taken from UK {\it Swift} Science Data Center
\footnote{http://www.swift.ac.uk/index.php} \citep{Eva+07, Eva+09}. 
We use the data observed by {\it Swift}/BAT \footnote{http://www.swift.ac.uk/burst\_analyser/} 
and XRT \footnote{http://www.swift.ac.uk/xrt\_curves/} to fit the light curve.
Our sample consists of short GRBs with at least three detection points by {\it Swift}/XRT. 
We also include several bursts with $T_{90}>2$ s in our sample, which are considered 
as short GRBs with extended emission detected by {\it Swift}/BAT \citep{GOWR13, Lien+16}.
Table 1 lists the sample of 65 short GRBs between January 2005 and June 2017, 
which corresponds to about a half of the entire short GRBs detected by {\it Swift}/BAT. 
Our sample overlaps with that in the previous studies of {\it Swift}/XRT-detected 
short GRBs, which were discussed in the context of the NS engine model \citep{Row+13, GOWR13, GOW14, Lu+15}.
For the bursts without known redshift, we use the averaged value of the measured-redshifts in our sample $z=$0.72, 
which is in agreement with the values reported in other works \citep[$<z>\sim0.5-0.8$; ][]{Row+13, D'Avanzo+14, B14, Lu+15}.

We show that after the prompt emission, 
the light curve consists of two components: the extended and plateau emission.
In order to identify the extended and plateau emission components, 
we adopt a phenomenological formula with two functions of a constant and subsequent power-law decay, 
\begin{eqnarray}\label{lightcurve}
L_{\rm iso}(t)&=&L_{\rm iso, EX}\left(1+\frac{t}{T_{\rm EX}}\right)^{-\alpha} \nonumber \\
& &+L_{\rm iso, PL}\left(1+\frac{t}{T_{\rm PL}}\right)^{-\alpha}, 
\end{eqnarray}
where $L_{\rm iso, EX}$, $L_{\rm iso, PL}$, $T_{\rm EX}$, and $T_{\rm PL}$ are 
the isotropic luminosities and durations of the extended and plateau emission, 
respectively \citep[see also ][]{Willingale+07}. 
In Equation (\ref{lightcurve}), the time after the {\it Swift}/BAT detection in the rest-frame is $t$, 
and the temporal index is $\alpha$.  
As a fiducial value, we use $\alpha=40/9$ implied by the BH engine model 
\citep[see Section \ref{sec:BHmodel} for details; ][]{KI15}, 
where the exact value does not alter our conclusions unless much small value $\alpha\lesssim2$ is assumed 
(see Section \ref{sec:discussion} for details). 
We define the extended emission as the emission with timescale $\lesssim10^3$ s, 
some of which are not detected by {\it Swift}/BAT. 
For a longer timescale component ($\gtrsim10^3$ s), we define it as the plateau emission. 
We compare the phenomenological formula with the observations and 
obtain the model parameters $L_{\rm iso, EX}, L_{\rm iso, PL}, T_{\rm EX}$ and $T_{\rm PL}$ for each burst.
We assume that the extended emission has to satisfy the condition $L_{\rm iso, EX}/L_{\rm iso, PL}\gtrsim10$, 
because a weak emission component is difficult to distinguish it from an X-ray flare. 
For the plateau emission, we require that there is at least one detection point whose luminosity is $>10$ times larger than 
that of the extended emission tail, $L_{\rm iso, EX}(t/T_{\rm EX})^{-\alpha}$ at $t>T_{\rm EX}$. 
Note that the identification of the extended and plateau emission is purely phenomenological. 

\section{RESULTS}
\label{sec:results}

\begin{figure*}
    \begin{tabular}{c}
      \begin{minipage}[t]{1.0\hsize}
        \centering
        \includegraphics[width=110mm, angle=270]{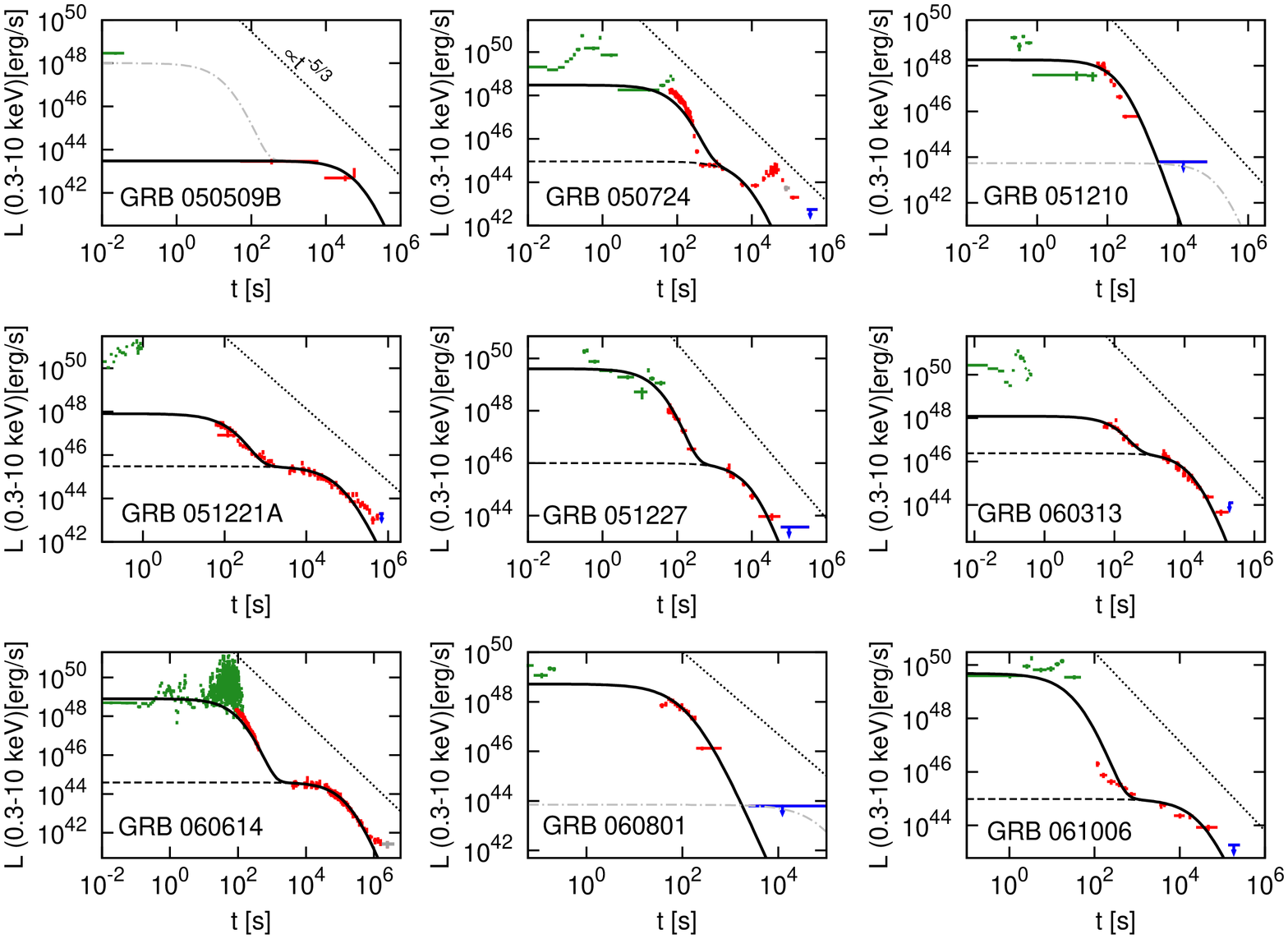}
      \end{minipage} \\
      \\
      \begin{minipage}[t]{1.0\hsize}
        \centering
        \includegraphics[width=110mm, angle=270]{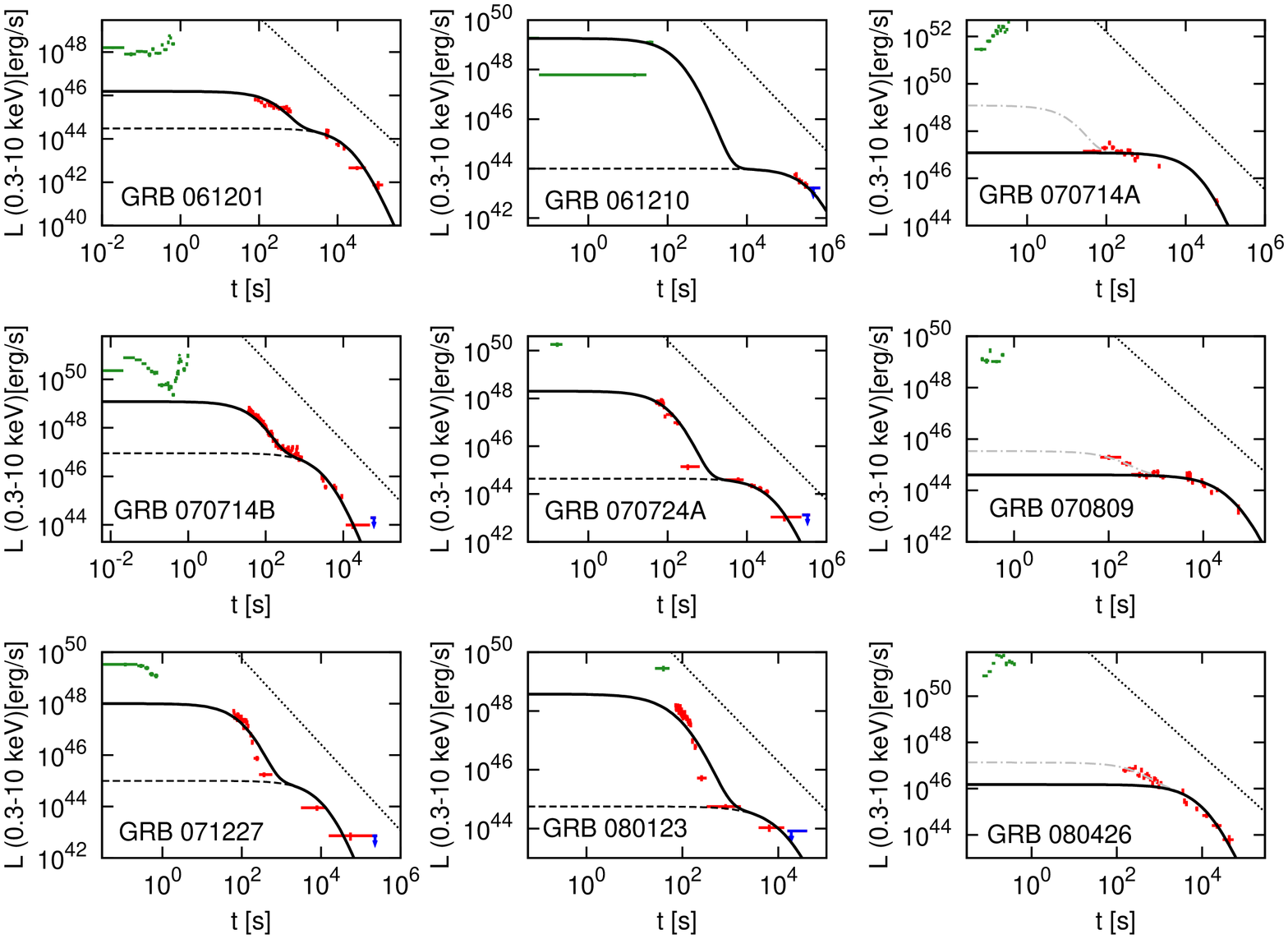}
        \caption{Model light curves for the extended and plateau emission of short GRBs 
as a function of the rest-frame time since the {\it Swift}/BAT trigger. 
Observational data is obtained from UK {\it Swift} Science Data Centre. 
Green and red points are the measured values by {\it Swift}/BAT and XRT, respectively. 
Gray points are the measured values with the exposure time less than $0.1$ times the duration error. 
Blue arrows are the upper limits on the luminosity. 
We do not fit the data at $t\le 2$ s, which is considered as the prompt emission. 
Gray dot-dashed curves show the possible components that were missed in the observations.}
        \label{fig:lightcurve}
      \end{minipage}
    \end{tabular}
\end{figure*}

\begin{figure*}
    \begin{tabular}{c}
      \begin{minipage}[t]{1.0\hsize}
        \centering
        \includegraphics[width=110mm, angle=270]{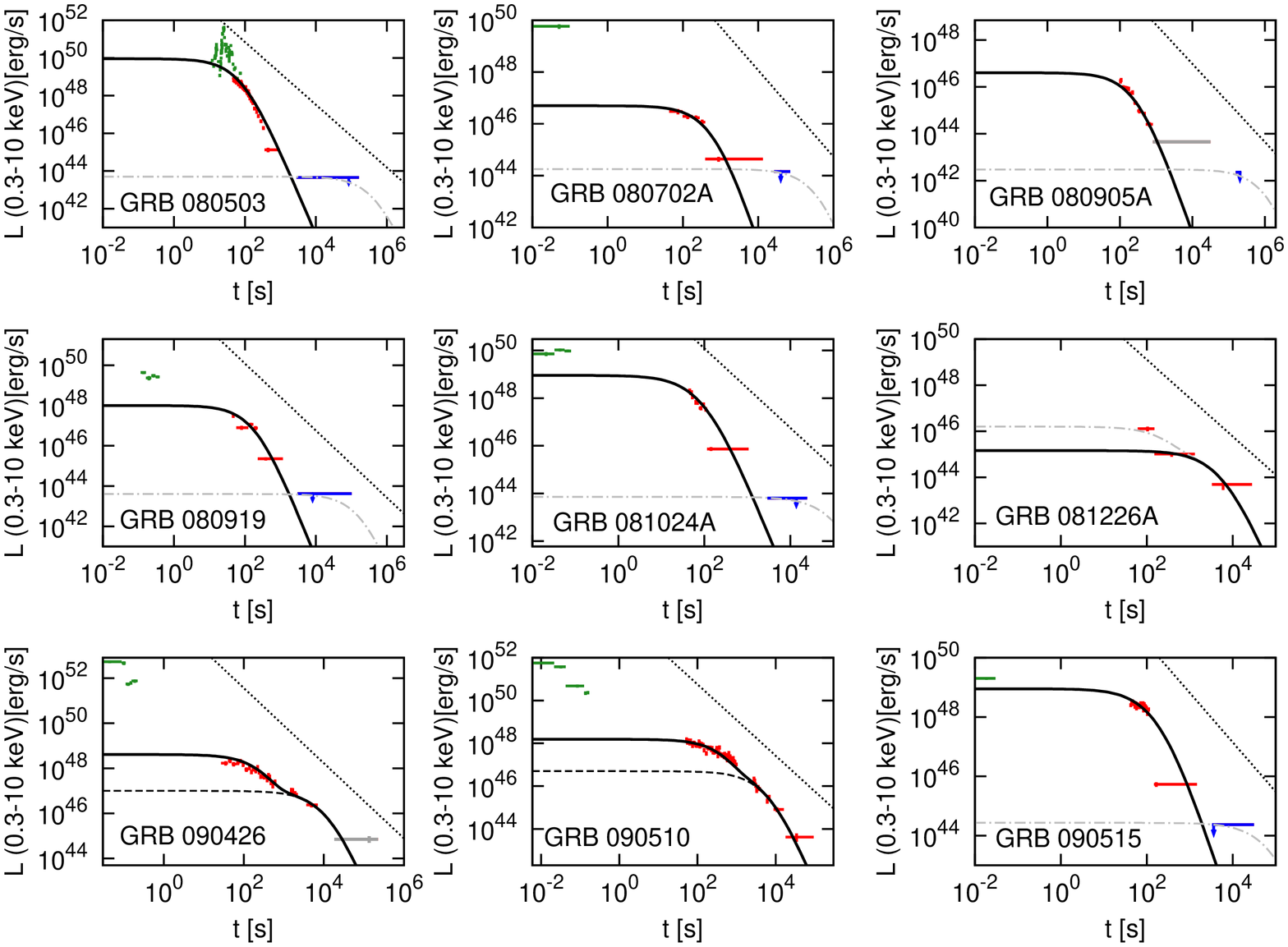}
      \end{minipage} \\
      \\
      \begin{minipage}[t]{1.0\hsize}
        \centering
        \includegraphics[width=110mm, angle=270]{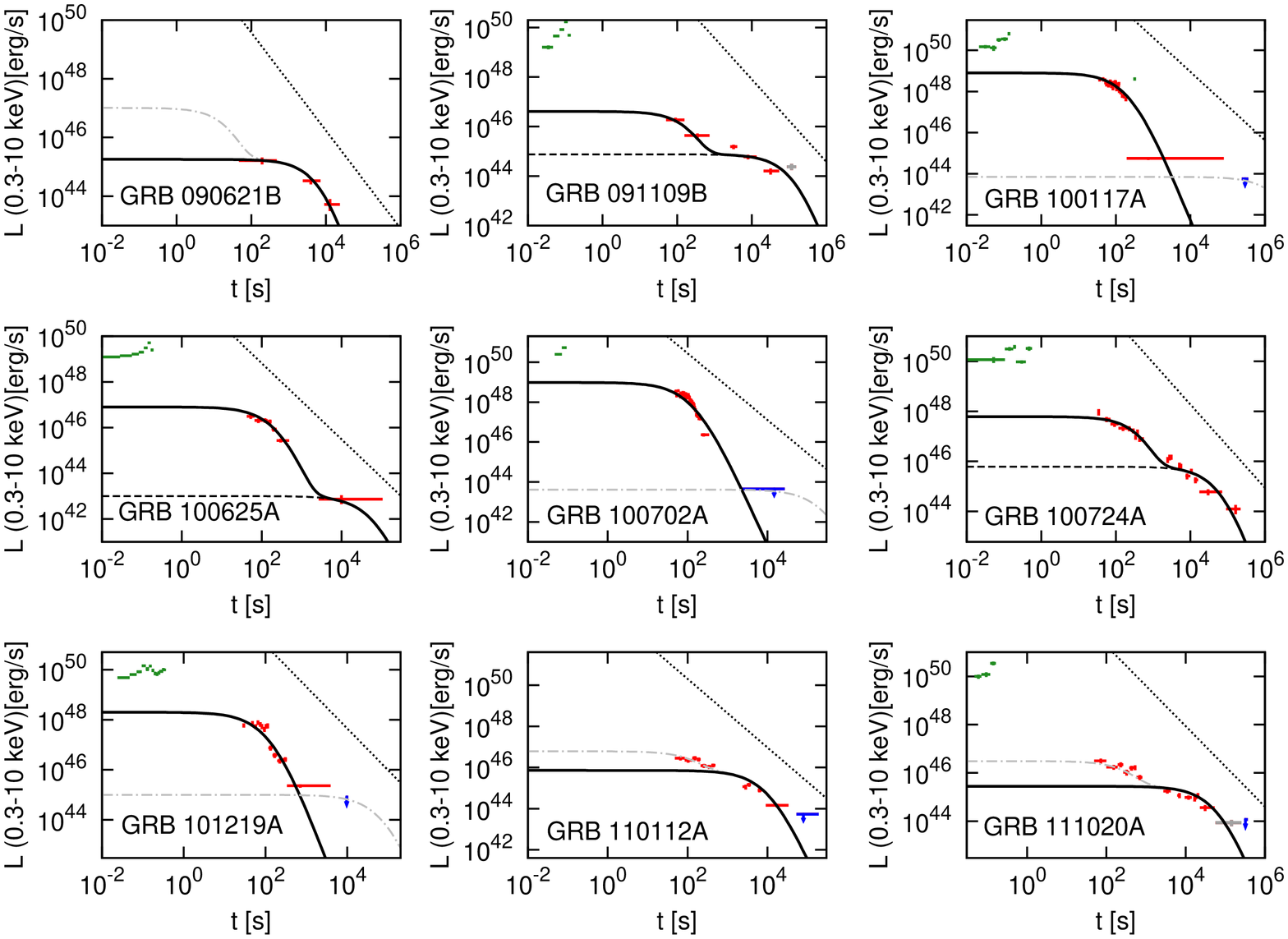}
         \center{Fig.1--- continued.}
      \end{minipage}
    \end{tabular}
\end{figure*}

\begin{figure*}
    \begin{tabular}{c}
      \begin{minipage}[t]{1.0\hsize}
        \centering
        \includegraphics[width=110mm, angle=270]{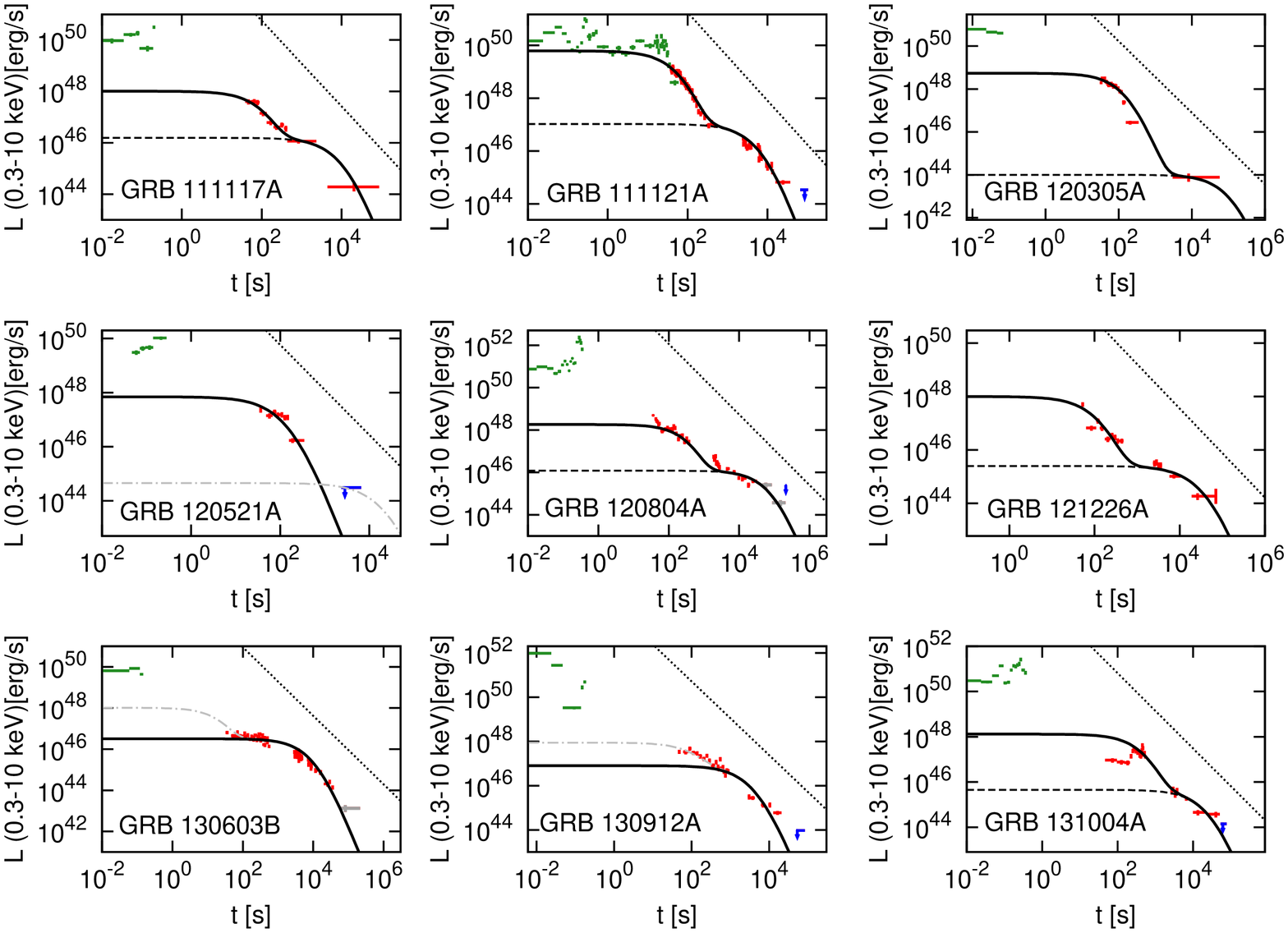}
      \end{minipage} \\
      \\
      \begin{minipage}[t]{1.0\hsize}
        \centering
        \includegraphics[width=110mm, angle=270]{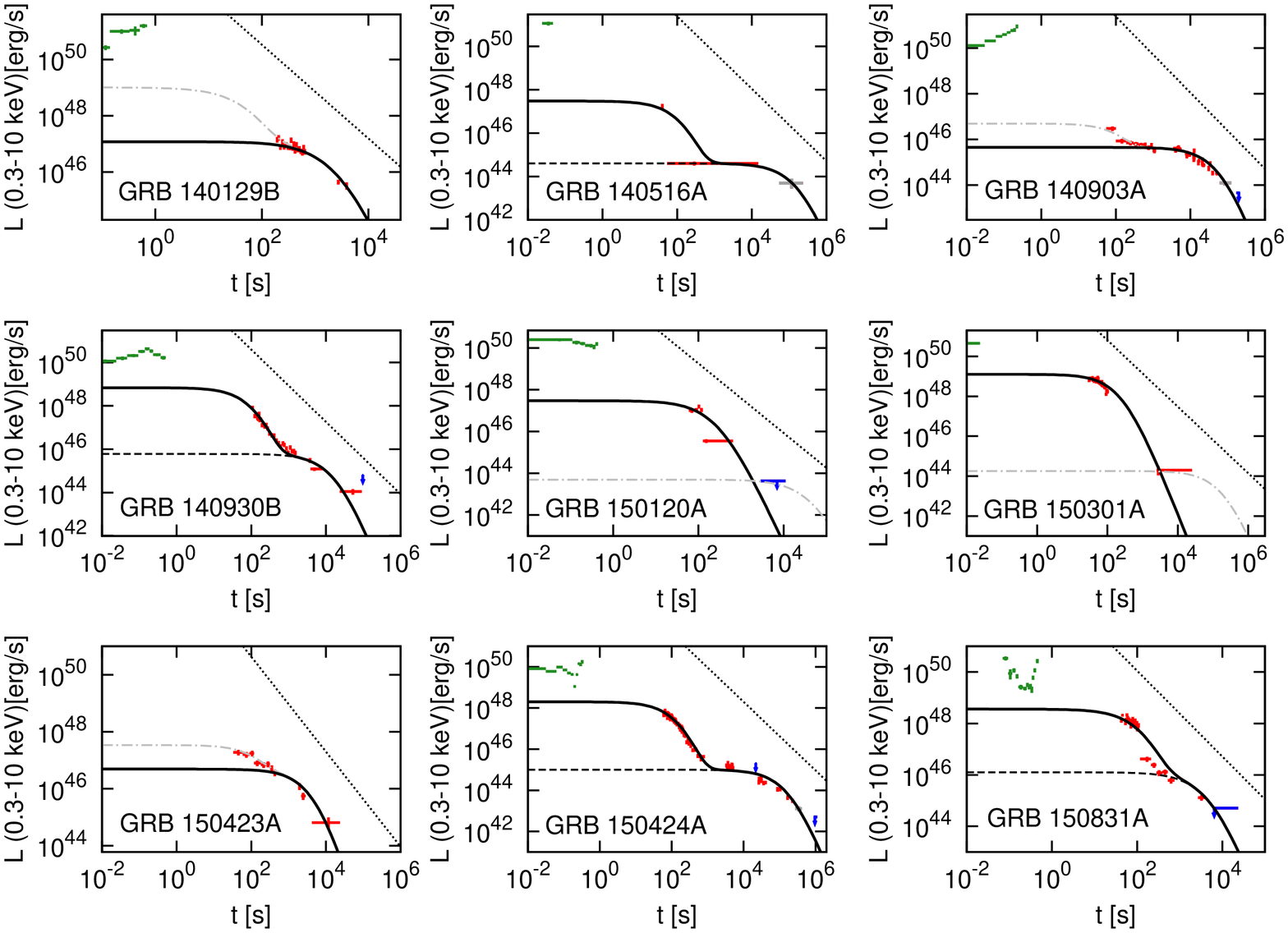}
         \center{Fig.1--- continued.}
      \end{minipage}
    \end{tabular}
\end{figure*}

\begin{figure*}
    \begin{tabular}{c}
      \begin{minipage}[t]{1.0\hsize}
        \centering
        \includegraphics[width=110mm, angle=270]{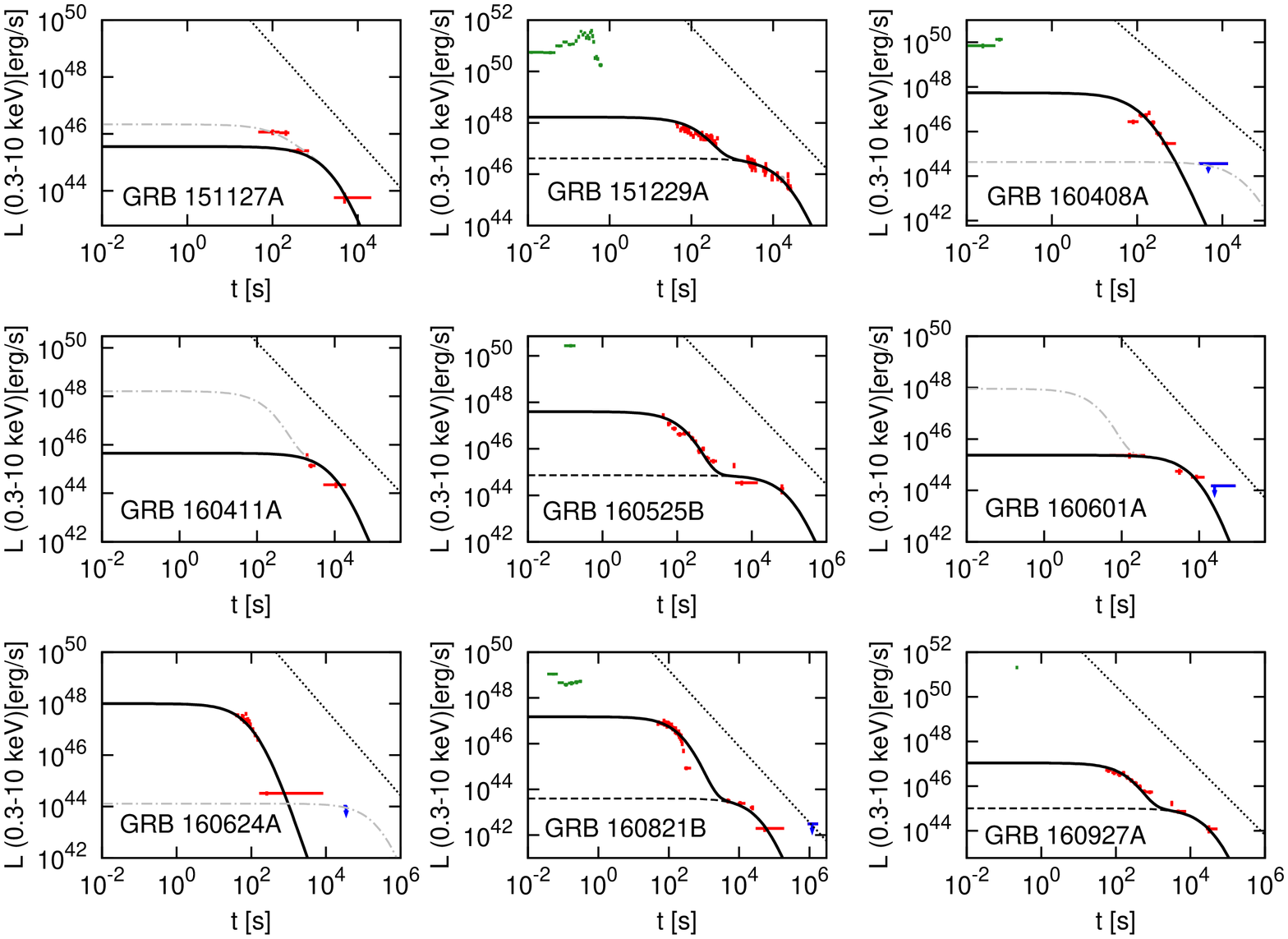}
      \end{minipage} \\
      \\
      \begin{minipage}[t]{1.0\hsize}
        \centering
        \includegraphics[width=36.7mm, angle=270]{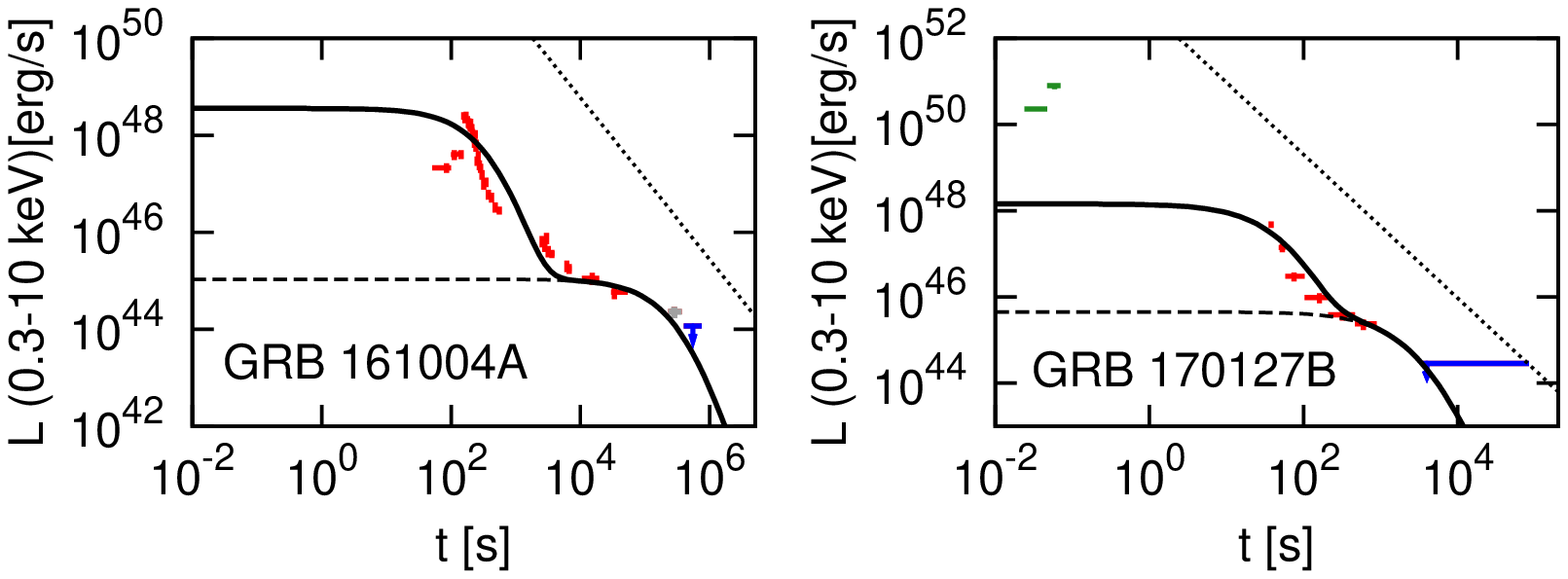}
         \center{Fig.1--- continued.}
      \end{minipage}
    \end{tabular}
\end{figure*}

Figure \ref{fig:lightcurve} shows the observational data points of all bursts in our sample. 
The phenomenological light curves 
in Equation (\ref{lightcurve}) with $\alpha=40/9$ are also shown as solid curves.  
The gray dot-dashed curves in Figure \ref{fig:lightcurve} denote the possible components that were missed in the observations. 
We fit the phenomenological light curve to the observational data by eye inspection, 
because the light curves often have some additional complex structures such as X-ray flares.
As shown in Figure \ref{fig:lightcurve}, the two-component light curves 
in Equation (\ref{lightcurve}) is consistent with {\it all} the observations. 
In some bursts, the decay of the extended emission is too sharp to fit the phenomenological model 
\citep[e.g., GRB 050724; ][]{Barthelmy+05}. 
We discuss this possible issue in Section \ref{sec:BHmodel}.

In Table 1, we list two components, 
the extended and plateau emission, seen in the observed light curves for each burst. 
The fraction of short GRBs with the extended emission in our sample detected by {\it Swift}/BAT and XRT is $49/65\sim0.75$.
This is almost the same value, $26/32\sim0.81$, also for the redshift-measured bursts. 
The number of short GRBs with the extended emission 
is about a half of the total {\it Swift}/BAT-detected short GRBs ($\sim100$ events).
These indicate that the {\it Swift}/BAT-detected short GRBs accompanying the extended emission is fairly common.
The number of bursts with the plateau emission is 49 for all sample and 26 for redshift-known sample,
which are (accidentally) the same values as the extended emission.
The number of short GRBs with both the extended and plateau emission is 33, which is a half of our sample.  
Therefore, the association of both components is also common for short GRBs.

 \begin{figure*}
  \begin{center}
   \includegraphics[width=160mm, angle=270]{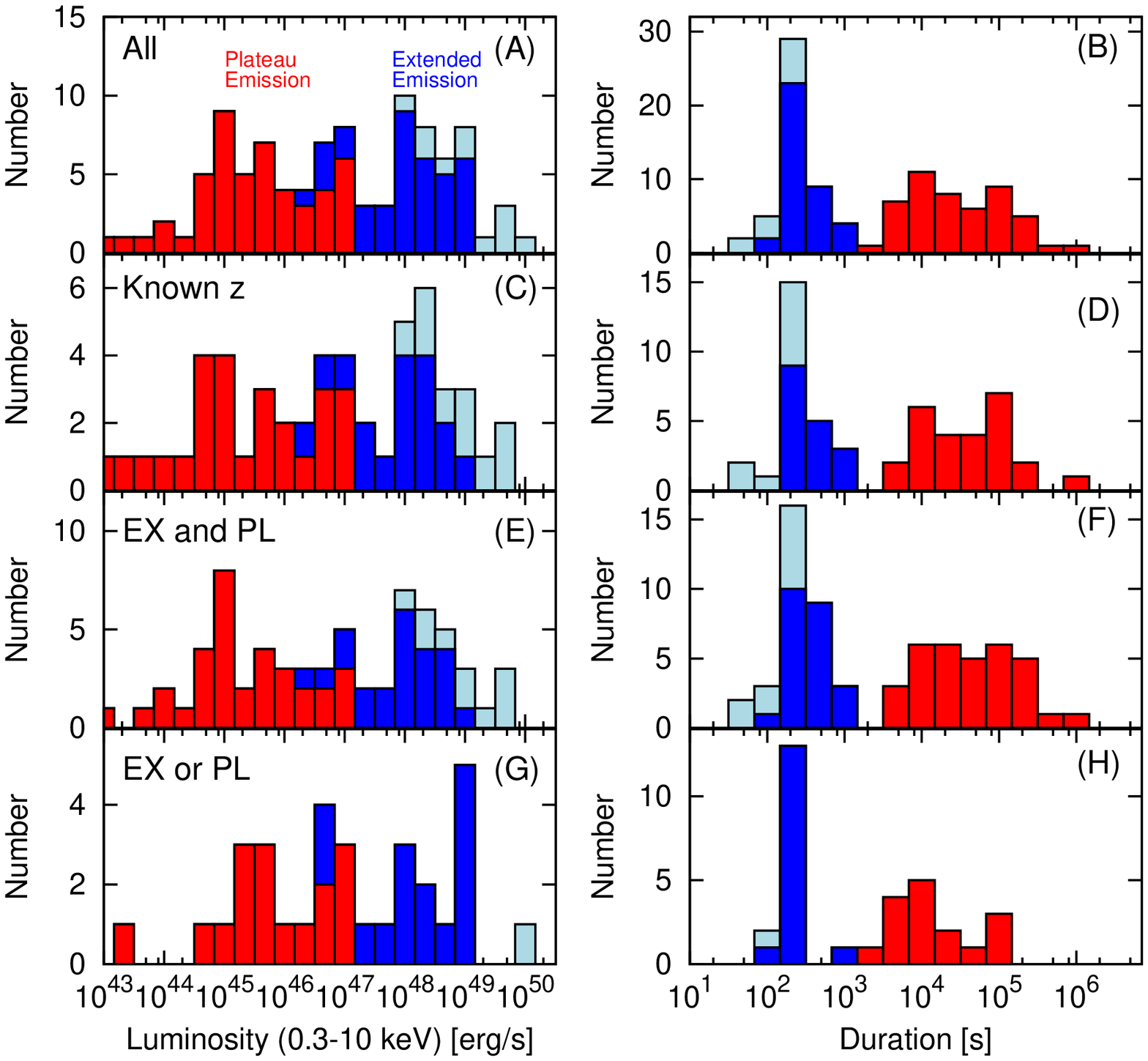}
   \caption{Distributions of the luminosity (0.3-10 keV; left)
     and duration (right) for the {\it Swift}/BAT-detected (light-blue) 
and BAT-non-detected extended emission (blue), and plateau emission (red) in the rest-frame. 
From top to bottom, panels show the distributions for all bursts (A and B), 
bursts with the measured-redshift (C and D), bursts with both the extended and plateau emission components (E and F), 
and bursts with single emission component (G and H). }
   \label{fig:histgram}
  \end{center}
 \end{figure*}

Figure \ref{fig:histgram} shows the luminosity (left panels) and duration distributions (right panels) for 
the {\it Swift}/BAT-detected (light-blue) and BAT-non-detected extended emission (blue), 
and plateau emission (red). 
The histograms show a hint of a bimodal distribution. 
To quantify the bimodality in the histograms, we perform the Hartigan's
Dip Test \citep{H85} using `diptest' CRAN package of the R software.
The null hypothesis of this test is that a distribution is a unimodal distribution. 
The null probabilities of the dip test of the luminosity distributions 
(Figure \ref{fig:histgram} (A), (C), (E) and (G)) are 0.404, 0.874, 0.341 
and 0.539 for the samples of all GRBs, redshift-known GRBs, GRBs with 
both the extended and plateau components and GRBs with the single component, respectively.
No statistically significant bimodality is evident in the luminosity distributions.
However, in the duration distributions (Figure \ref{fig:histgram} (B), (D), (F) and (H)),
the null probabilities are $3.67 \times 10^{-3}$, $3.91 \times 10^{-3}$,
$4.10 \times 10^{-3}$ and $1.73 \times 10^{-2}$ for the samples of all GRBs,
redshift-known GRBs, GRBs with both the extended and plateau components
and GRBs with the single component, respectively.  
Therefore, the duration distributions reject a unimodality, 
and prefer a bimodal distribution.
Based on this Dip Test, the extended and plateau emission are very likely distinct populations.
Namely, there are two distinct long-lasting components following the prompt emission in short GRBs. 
This is the first indication that two distinct long-lasting components are ubiquitous 
in short GRB light curves as far as we know (see the cluster analysis in Figure \ref{fig:L-T} for stronger evidence). 
A normal afterglow cannot explain the plateau-like temporal evolution in the light curve \citep[e.g., ][]{SPN98}.
The presence of the long-lasting components means that there are at least two activity phases 
related to the central engine in addition to the prompt emission. 
Our findings would not only serve as a key to the final understanding of the short GRBs, 
but also provide appropriate strategies to detect short GRBs as 
an EM counterpart to a GW source (Section \ref{sec:detectability}).

Figures \ref{fig:histgram} (A) and (B) show the distributions for all bursts in our sample.
From Figure \ref{fig:histgram} (A), 
the luminosity range of the extended emission is $10^{47}$ erg s$^{-1}\lesssim L_{\rm iso, EX} \lesssim10^{50}$ erg s$^{-1}$, 
which is somewhat broader than that of the extended emission detected by {\it Swift}/BAT 
\citep[$10^{48}$ erg s$^{-1}\lesssim L_{\rm iso, EX}\lesssim10^{50}$ erg s$^{-1}$; e.g., ][]{GOWR13, GOW14}.
The duration distribution of the extended emission is concentrated in $T_{\rm EX}\sim200$ s (Figure \ref{fig:histgram} B).
This narrowness of the $T_{\rm EX}$ distribution might be the intrinsic properties, 
or the observational bias that we are missing the extended emission with shorter duration due to time lag 
between the BAT triggering time and the observational starting time of {\it Swift}/XRT, $\sim60-100$ s.
The luminosity and duration of the {\it Swift}/BAT-detected and
non-detected extended emission are continuously distributed, 
so that both populations would be the same component.
The luminosity range of the plateau emission is $10^{43}$ erg s$^{-1}\lesssim L_{\rm iso, PL}\lesssim10^{47}$ erg s$^{-1}$.
The duration of the plateau emission is $T_{\rm PL}\sim10^4-10^5$ s. 
For comparison, Figures \ref{fig:histgram} (C) and (D) show only the distributions for redshift-known bursts.
There is no clear difference between the distributions for all and redshift-known bursts. 

In Figures \ref{fig:histgram} (E-H), we show the luminosity and duration distributions 
to compare between bursts with both components (panels E and F) and with single component (panels G and H).
For the plateau emission, 
there seems to be some differences that the luminosity and duration of the bursts 
without the detectable extended emission are respectively high and short on average. 
A possible bias is that the exposure time of {\it Swift}/XRT is set to be too short 
to detect the low-luminosity and long-duration plateau emission for bursts without detectable extended emission.  

\begin{figure*}
  \begin{center}
   \includegraphics[width=80mm, angle=270]{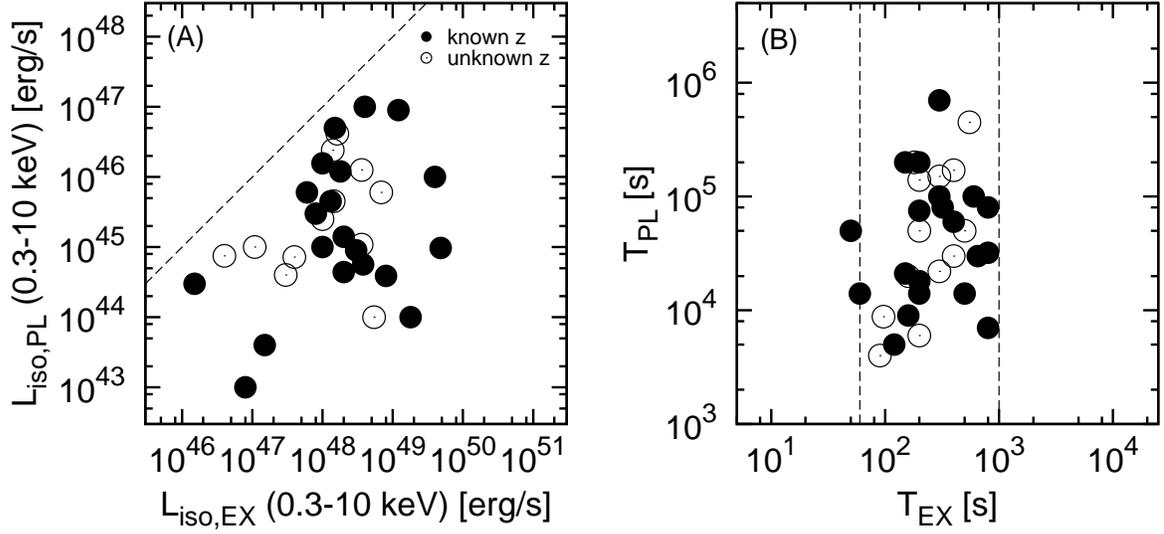}
   \caption{$L_{\rm iso, EX}$ vs.$L_{\rm iso, PL}$ (0.3-10 keV; A) and $T_{\rm EX}$ vs. $T_{\rm PL}$ plots (B) 
 for bursts with both the extended and plateau emission components in the rest-frame. 
 Filled symbols are bursts with known redshift, and open symbols are bursts without measured redshift.
 Dashed line in panel (A) is the criterion of the extended emission in this paper, $L_{\rm iso, PL}=0.1L_{\rm iso, EX}$.
 Two vertical dashed lines in panel (B) show the typical starting time of the {\it Swift}/XRT observations ($\sim60$ s) 
 and the maximum timescale of the extended emission in our criterion ($10^3$ s). }
   \label{fig:L-L_T-T}
  \end{center}
 \end{figure*}

Figure \ref{fig:L-L_T-T} shows the luminosity-luminosity (A) and duration-duration plots (B) 
for the extended and plateau emission 
obtained from the fitting of the observed light curves.
Plotted data are only short GRBs with both the extended and plateau emission. 
Because of our criterion for the extended emission ($L_{\rm iso, EX}/L_{\rm iso, PL}>10$), 
there is no event at the upper left region from the dashed line, $L_{\rm iso, PL}=0.1L_{\rm iso, EX}$, 
in Figure \ref{fig:L-L_T-T} (A).
Although a weak positive trend may be seen in Figure \ref{fig:L-L_T-T} (A), 
there is a significant scatter in the distribution. 
For the durations, no significant correlation is seen in Figure \ref{fig:L-L_T-T} (B). 
Therefore, it is difficult to predict the luminosity and duration of the plateau emission
with an accuracy of a factor of a few from those of the extended emission.

\begin{figure*}
  \begin{center}
   \includegraphics[width=100mm]{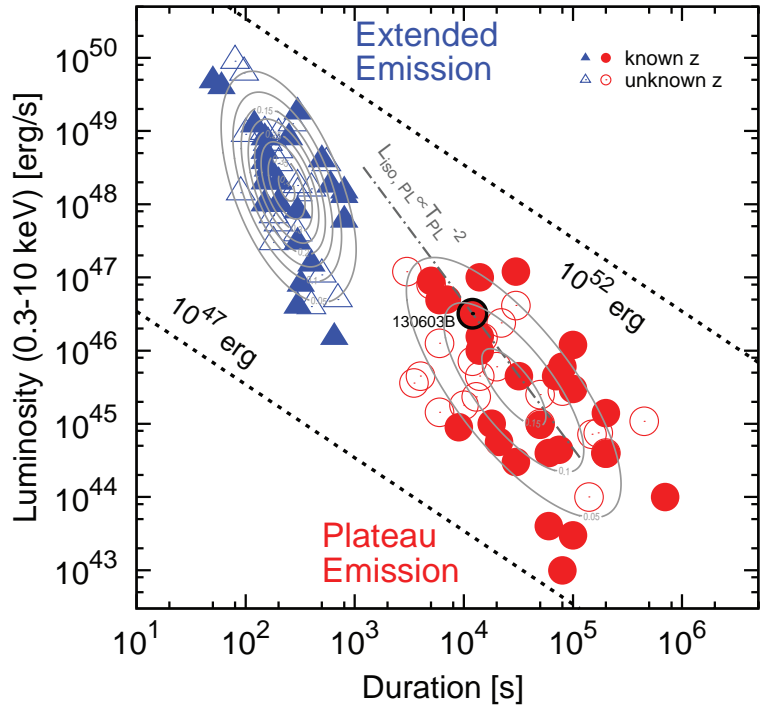}
   \caption{Luminosity (0.3-10 keV) vs. duration plot of
     the extended (blue) and plateau emission (red) in the rest-frame. 
Filled symbols are bursts with known redshift, and open symbols are bursts without measured redshift.
The contours show the density estimate results in the cluster analysis. 
The null hypothesis that the data are represented by a unimodal distribution is rejected at $>5.2\sigma$.
Gray dot-dashed line shows the relation $L_{\rm iso, PL}\propto T_{\rm PL}^{-2}$ (see Section \ref{sec:detectability}).}
   \label{fig:L-T}
  \end{center}
 \end{figure*}

Figure \ref{fig:L-T} shows the luminosity - duration plot of the extended and plateau emission
for all bursts of our sample. 
A notable feature is that 
two parameter regions of the extended and plateau emission in Figure \ref{fig:L-T} are clearly separated. 
To quantify this finding, we apply model-based cluster analysis to the data using `mclust' CRAN package 
\citep{FR02} of the R software.  
The luminosity - duration data can best be classified into two groups, 
which correspond to the clusters of the extended emission (blue of Figure \ref{fig:L-T}) 
and the plateau emission (red of Figure \ref{fig:L-T}).
The significance of this bimodality is evaluated by the bootstrap approach 
using `mclustBootstrapLRT' program which is a part of the `mclust' package. 
Based on this analysis, the null hypothesis that the data are represented 
by a unimodal distribution is rejected at $>$5.2$\sigma$.
Therefore, our result strongly supports that the extended and plateau emission 
are distinct components from each other. 
There is a general trend that the plateau emission with lower luminosity has longer duration (Figure \ref{fig:L-T}), 
although there is a large scatter. 
From the light curve formula in Equation (\ref{lightcurve}), 
the isotropic energy for each component $i(=$EX or PL) in 0.3-10 keV is 
\begin{eqnarray}
E_{\rm iso, i}=\frac{L_{\rm iso, i}T_{\rm i}}{\alpha-1}~~~(\alpha>1).
\end{eqnarray}
Using the value $\alpha=40/9$, the isotropic energy is $E_{\rm iso, i}=(9/31)L_{\rm iso, i}T_{\rm i}$. 
From Figure \ref{fig:L-T}, 
the ranges of the isotropic energies of the extended and plateau emission are 
$10^{48}~{\rm erg}\lesssim E_{\rm iso, EX}\lesssim 10^{51}~{\rm erg}$ and 
$10^{47}~{\rm erg}\lesssim E_{\rm iso, PL}\lesssim 10^{51}~{\rm erg}$, respectively. 

\begin{figure*}
  \begin{center}
   \includegraphics[width=100mm, angle=270]{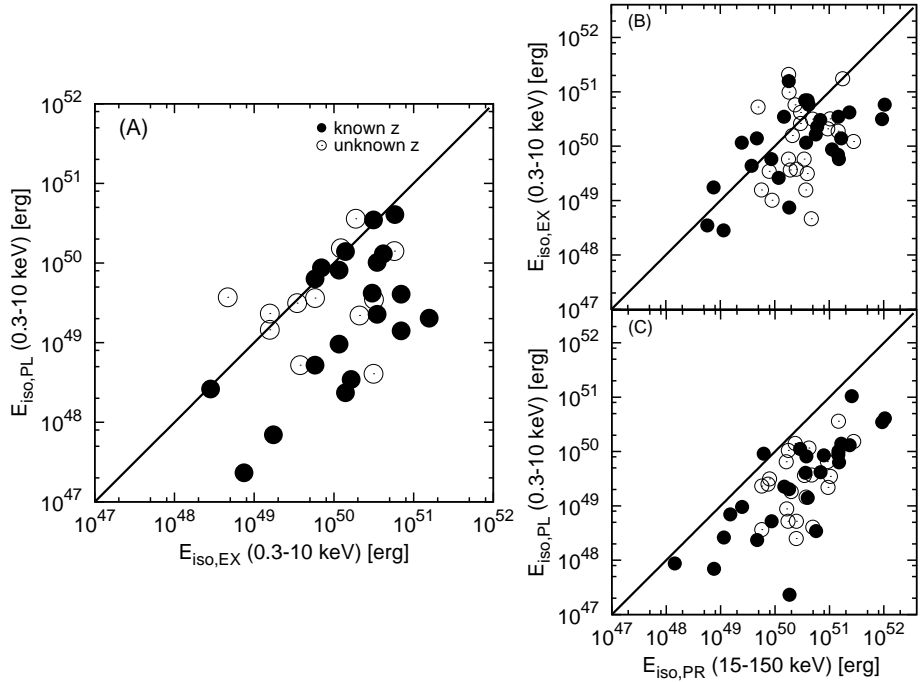}
   \caption{Plots of the isotropic energies $E_{\rm iso, PL}$ (0.3-10 keV), $E_{\rm iso, EX}$ (0.3-10 keV) 
and $E_{\rm iso, PR}$ (15-150 keV). 
Filled circles are bursts with known redshift, 
and open circles are bursts without measured-redshift.}
   \label{fig:EE-PL}
  \end{center}
 \end{figure*}

Figure \ref{fig:EE-PL} (A) plots the energies $E_{\rm iso, EX}$ and $E_{\rm iso, PL}$ 
for short GRBs with both components.
The ratio of the energy is $E_{\rm iso, EX}/E_{\rm iso, PL}\sim1-100$. 
Figure \ref{fig:EE-PL} (A) indicates that there is a possible positive trend between two energies 
$E_{\rm iso, EX}$ and $E_{\rm iso, PL}$ for all and redshift-measured bursts (filled circles). 

Figures \ref{fig:EE-PL} (B) and (C) show the isotropic energy plots of $E_{\rm iso, EX}$ and $E_{\rm iso, PL}$ 
relative to that of the prompt emission, $E_{\rm iso, PR}$.
We use the {\it Swift}/BAT fluence (15-150 keV) taken from the {\it Swift} GRB Table 
\footnote{https://swift.gsfc.nasa.gov/archive/grb\_table/} to calculate $E_{\rm iso, PR}$. 
For the bursts with the {\it Swift}/BAT-detected extended emission, 
we use the fluence of only a short pulse in the BAT energy band for $E_{\rm iso, PR}$ 
\citep[Data are provided by A. Lien for GRBs listed in Table 3 of ][]{Lien+16}. 
For GRB 060614, we use the fluence of the short pulse from \citet{Gehrels+06}.
For GRB 080123, since the extended emission in the {\it Swift}/BAT band was weak \citep{Lien+16}, 
we use the total fluence from the {\it Swift} GRB Table as the fluence of the prompt emission. 
Note that the observed peak energy of the prompt emission of short GRBs $E_{\rm peak}^{\rm obs}$ could be higher than 150 keV, 
and the photon index $\alpha_{\rm p}$ is typically $\sim-1$ \citep[e.g., ][]{Lien+16}. 
In the case of $E_{\rm peak}^{\rm obs}>150$ keV, 
the bolometric isotropic energy could be $\sim(E_{\rm peak}^{\rm obs}/150~{\rm keV})^{2+\alpha_{\rm p}}$
times higher than that in 15-150 keV band. 

Figure \ref{fig:EE-PL} (B) shows that the energies of the prompt and extended emission are almost comparable, 
$E_{\rm iso, PR}\sim E_{\rm iso, EX}$.
The short GRBs with the {\it Swift}/BAT-detected extended emission have a similar trend \citep[e.g., ][]{Per+09}. 
This also supports that both the extended emission detected by {\it Swift}/BAT and XRT are the same component. 
For the plateau emission, although the isotropic energy is on average $E_{\rm iso,PL}\sim0.1E_{\rm iso, PR}$, 
a significant fraction of 
bursts has $E_{\rm iso, PL}$ comparable to $E_{\rm iso, PR}$ as shown in Figure \ref{fig:EE-PL} (C). 
This would suggest that the plateau emission is also produced by the central engine activities. 
The energy $E_{\rm iso, PL}$ also seems to have a positive trend with $E_{\rm iso, PR}$.

\section{Detectability as Electromagnetic Counterparts to Gravitational Wave Sources}
\label{sec:detectability}

\begin{figure*}
  \begin{center}
   \includegraphics[width=160mm, angle=270]{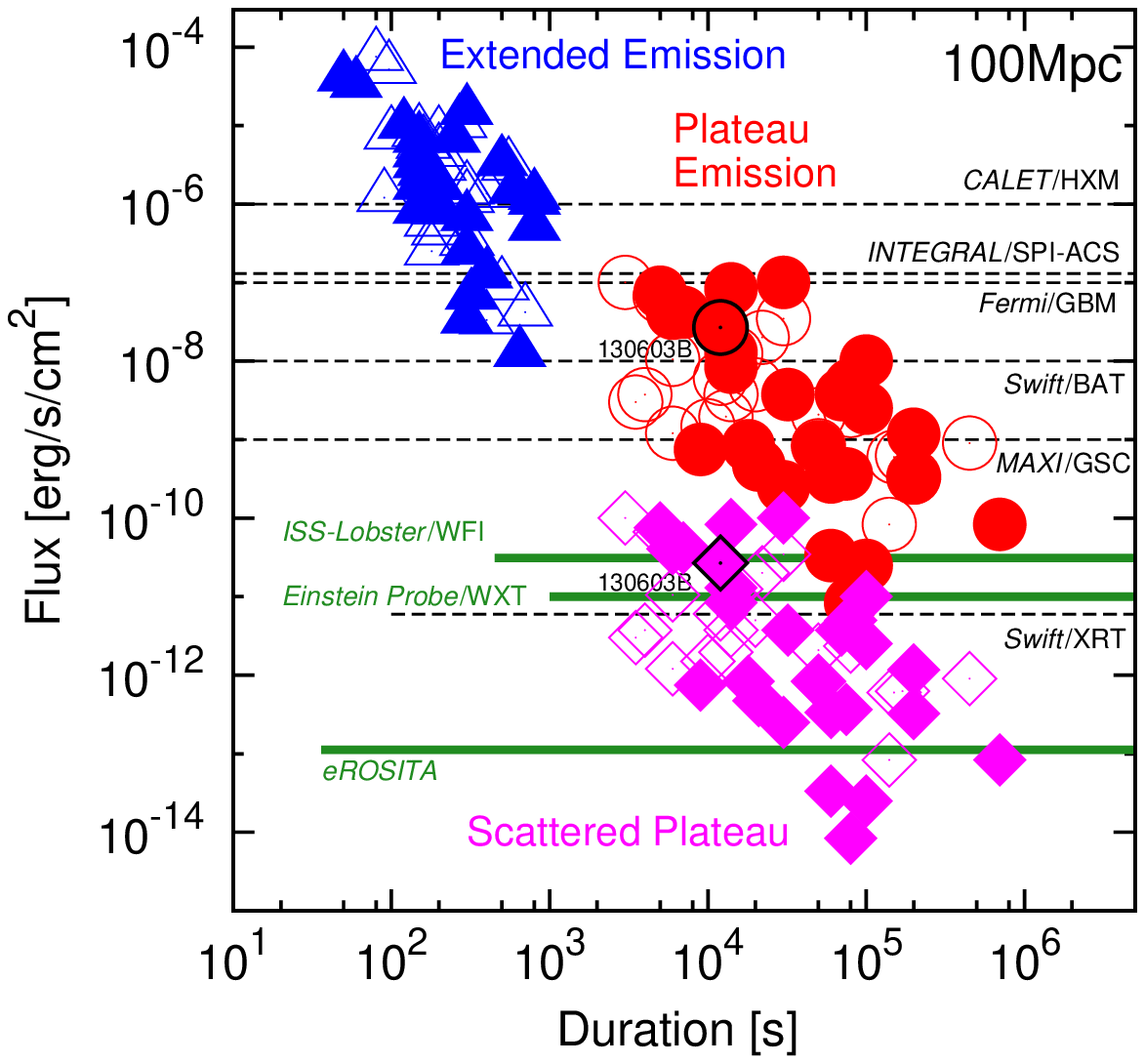}
   \caption{Flux-duration plot of extended, plateau and scattered emission components for all short GRBs with distance 100 Mpc. 
Black dashed lines show the sensitivity limits of {\it CALET}/HXM \citep[integration time 1 s; ][]{Adriani+16}, 
{\it INTEGRAL}/SPI-ACS \citep[integration time 1 s; ][]{Savchenko+16}, 
{\it Fermi}/GBM \citep[integration time 1 s; ][]{Bhat+16}, 
{\it Swift}/BAT \citep[integration time 1 s; ][]{Krimm+13}, 
{\it Swift}/XRT \citep[integration time 100 s; ][]{Kanner+12}, and 
{\it MAXI}/GSC \citep[integration time corresponding to a single pass; ][]{Sugizaki+11}. 
Green solid lines show the sensitivity limits of planned detectors, 
{\it ISS-Lobster}/WFI \citep[integration time 450 s; ][]{Camp+13}, 
{\it Einstein Probe}/WXT \citep[integration time 1000 s; ][]{Yuan+15}, 
and {\it eROSITA} \citep[integration time corresponding to a single survey pass; ][]{Mer+12}. 
}
   \label{fig:detectability}
  \end{center}
 \end{figure*}

The leading model of short GRBs is an NS binary merger \citep[e.g., ][]{NPP92}. 
The NS binary merger is accompanied with strong GW emission, which can be detected by current GW detectors. 
A simultaneous detection of GW and EM emission would maximize the available information 
from this spectacular event \citep[e.g., ][]{MB12}.

Current $\gamma$-ray detectors such as {\it Swift}/BAT can detect the prompt emission 
if the short GRB occurs within the GW detection horizon ($\sim100$ Mpc) and the GRB jet points to us. 
Since the duration of the prompt emission is $\lesssim1$ s, the detection probability is 
mainly determined by the field-of-view of the detectors. 

The long-lasting components, the extended and plateau emission, are also bright enough for
detections if a short GRB occurs within the GW detection horizon.
By virtue of the long duration, the follow-up observations are possible after receiving the GW detection alert 
\citep[within $\lesssim10^2$ s; ][]{Singer+14, CH15, SP16}. 
Especially, since the duration of the plateau emission is longer than the orbital period of all-sky survey detectors, 
the detection probability is much higher than that determined by the ratio of the field-of-view to the all-sky.

Nearly isotropic emission is also expected. 
A significant fraction of the plateau emission could be scattered 
to a wide solid angle by the merger ejecta \citep{KIN15}. 
During the plateau activity timescale ($\sim10^4$ s), 
the radius of the plateau emission region is smaller than 
that of the expanding merger ejecta. 
The optical depth for the Thomson cross section is typically larger than unity during the plateau phase. 
Since the Lorentz factor is low ($\Gamma\sim10$) inside the jet due to the cocoon confinement \citep{BNPS11, Nagakura+14}, 
the relativistic beaming angle is larger than the jet opening angle $\theta_{\rm j}$. 
Then, the emitted photons with angle $\gtrsim\theta_{\rm j}$ relative to the jet axis 
are scattered to a large angle by the surrounding non-relativistic ejecta \citep{KIN15}.
Since the collimated emission is scattered to an isotropic distribution, 
the luminosity of the scattered component is 
\begin{eqnarray}\label{scatter}
L_{\rm scatter}\sim10^{-3}(\theta_{\rm j}^2/10^{-3})L_{\rm iso, PL}. 
\end{eqnarray}

Here, we investigate the detectability of on-axis extended and plateau emission 
taking into account their luminosity and duration distributions.
We also consider the detectability of the scattered plateau emission.
Figure \ref{fig:detectability} shows the energy flux and duration of the extended and plateau emission 
if the short GRBs in our sample occur at 100 Mpc, which is approximately the detection horizon of a binary NS merger
for the current GW detectors \citep{Abbott+16d}. 
We use Equation (\ref{scatter}) to estimate the flux of the scattered plateau emission \citep{KIN15}.
In Figure \ref{fig:detectability}, we also plot the flux sensitivities of the current and planned X-ray detectors. 
Since the typical photon index of the extended emission is $\sim-2$ \citep{Lien+16} and a similar value 
within the uncertainty for the plateau emission (according to {\it Swift}/XRT GRB light curve repository 
\footnote{http://www.swift.ac.uk/xrt\_curves/}), 
we neglect the difference of the energy bands for each detector to consider the energy flux. 

From Figure \ref{fig:detectability}, 
the energy flux distributions of the extended and plateau emission are $\sim10^{-7}-10^{-4}$ erg cm$^{-2}$ s$^{-1}$
and $\sim10^{-11}-10^{-7}$ erg cm$^{-2}$ s$^{-1}$, respectively. 
The monitoring observations by {\it CALET}/HXM, {\it Fermi}/GBM, {\it INTEGRAL}/SPI-ACS, and {\it Swift}/BAT 
can detect most of the extended emission population if the line of sight is within the jet opening angle. 
{\it Swift}/BAT and {\it MAXI}/GSC can detect some bright plateau emission. 
Note that {\it MAXI} scans the entire sky every 92-minute orbital period, 
which is comparable to or shorter than the duration of the plateau emission. 
Although {\it eROSITA} can detect most of the scattered plateau emission, 
its small field-of-view (0.833 deg$^2$) makes it difficult to detect the X-ray signals with GW simultaneously.

In the follow-up observations of the GW sources, 
{\it Swift}/XRT with 100 s integration time \citep{Kanner+12}, 
future {\it ISS-Lobster}/WFI with 450 s integration time \citep{Camp+13}, and future {\it Einstein Probe}/WXT 
with 1000 s integration time \citep{Yuan+15} can detect the plateau emission. 
These detectors can also detect the scattered plateau emission in the bright population including GRB 130603B 
(Figure \ref{fig:detectability}). 

The long-lasting activities of the central engine could significantly contribute to 
the heating of the merger ejecta \citep[engine-powered macronova; ][]{K05, YZG13, MP14, KIT15, KIN15}.
The emission from the heated ejecta is observed as a macronova 
and has been discussed as a promising EM counterpart to the
NS binary merger \citep[e.g., ][]{LP98, K05, Metzger+10, KIT15, KIN15, KIN16}. 
The observed peak luminosity depends on the internal energy in the ejecta at the time when the diffusion timescale
of photons in the ejecta becomes comparable to the dynamical timescale.
If the energy injection timescale is earlier than the peak phase of the macronova, 
the internal energy in the ejecta decreases due to the adiabatic cooling.  
Then, the  plateau emission is more effective for heating than the extended emission 
\citep{KIN15}. 
The internal energy in the heated ejecta after the energy injection ($t>T_{\rm PL}$) is 
$\sim E_{\rm int}(t/T_{\rm PL})^{-1}\propto L_{\rm iso, PL}T_{\rm PL}^2$, 
where the total injected energy is determined by the radiative efficiency $\eta$, 
the jet opening angle $\theta_{\rm j}$, 
the observed isotropic luminosity $L_{\rm iso, PL}$, and the duration $T_{\rm PL}$, as
$E_{\rm int}=[(\theta_{\rm j}^2/2)/\eta]L_{\rm iso, PL}T_{\rm PL}$. 
Note that the value of $L_{\rm iso, PL}T_{\rm PL}^2$ for GRB 130603B 
($L_{\rm iso, PL}\sim4\times10^{46}$ erg s$^{-1}$, $T_{\rm PL}\sim10^4$ s), 
which was first reported as a macronova event \citep{Tanvir+13, BFC13}, 
lies in the median of the distribution of $L_{\rm iso, PL}T_{\rm PL}^2$ (gray dot-dashed line in Figure \ref{fig:L-T}).
Thus the peak luminosity of the macronova associated with GRB 130603B is a typical value of the engine-powered macronova, 
implying that the dominant energy source could be the central engine not the radioactivity of $r$-process elements. 

\section{Implications for BH Engine Models}
\label{sec:BHmodel}

 \begin{figure*}
  \begin{center}
   \includegraphics[width=120mm, angle=270]{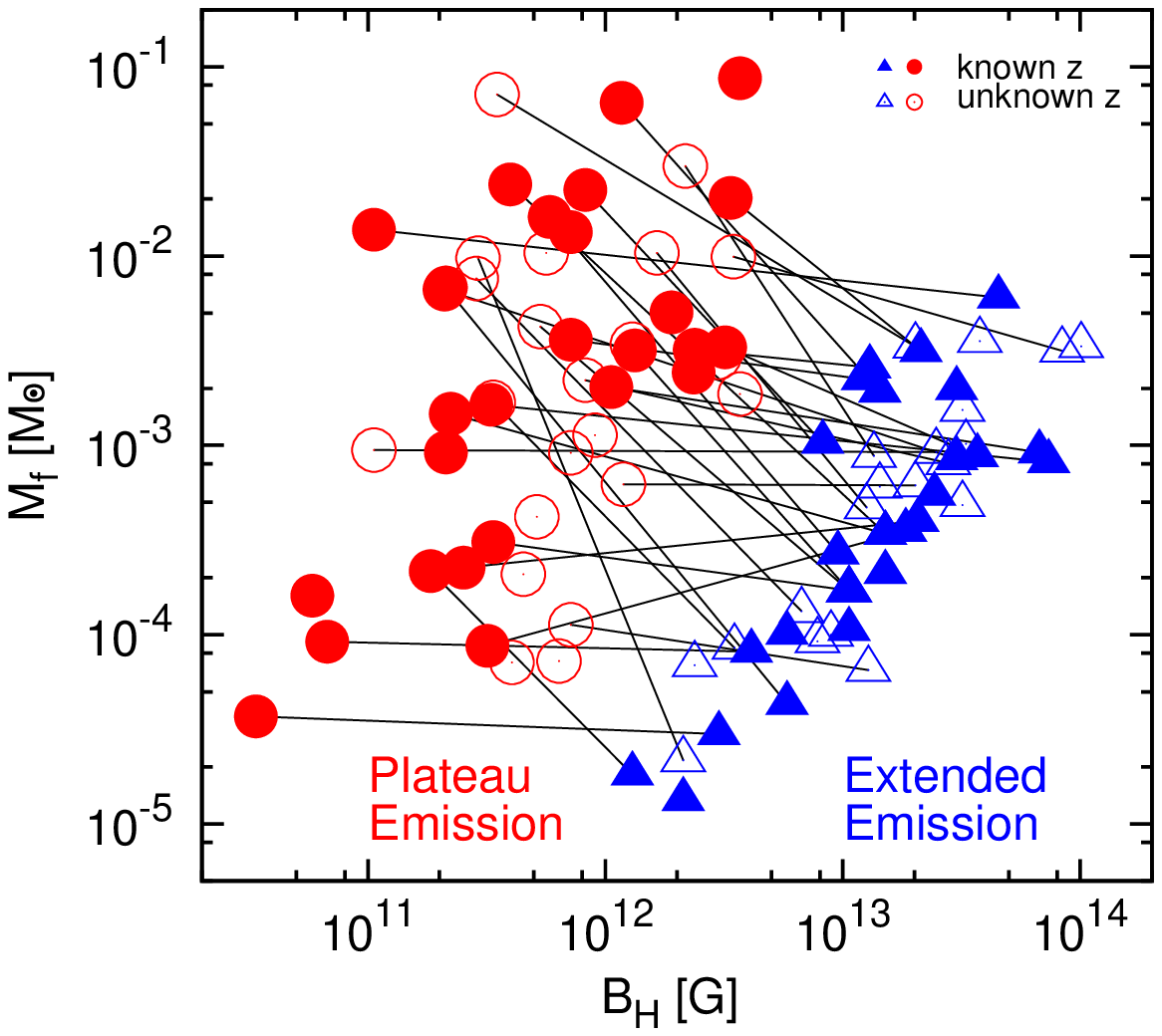}
   \caption{Estimated total mass of the fallback matter $M_{\rm f}$ and magnetic field $B_{\rm H}$ 
for extended (blue triangles) and plateau emission (red circles). 
Filled symbols are events with known redshift, and open symbols are events without measured redshift. 
Solid lines connect each event of short GRBs.}
   \label{BHmodel}
  \end{center}
 \end{figure*}

In Section 3, we show that the light curve of most short GRBs consists of bimodal long-lasting components 
following the prompt emission.
The plateau-like evolution ($L\propto t^0$) is difficult to explain by the normal afterglow model 
\citep[e.g., ][]{SPN98}. 
For a single plateau model such as the spin-down of highly magnetized NSs 
\citep[e.g., ][]{U92, ZM01, GF06, MQT08, BMTQ12, L13} and BHs \citep[e.g., ][]{BP11, Nakamura+14}, 
additional mechanisms to produce another component are required \citep{GOW14, GWGO17}.

The phenomenological light curve in Equation (\ref{lightcurve}) is motivated by the BH engine model \citep{KI15}. 
An NS-NS or BH-NS merger leaves a BH with a surrounding disk and merger ejecta \citep[e.g., ][]{Hot+13, Kyu+15}. 
In this model, a relativistic jet is launched via Blandford-Znajek (BZ) process \citep{BZ77} from the BH. 
For the BH with a mass $M_{\rm BH}$, a spin parameter $a=Jc/GM_{\rm BH}$, an angular frequency 
$\Omega_{\rm H}=ac/(2M_{\rm BH}r_{\rm H})$, and a magnetic flux $\Psi_{\rm BH}\sim\pi r_{\rm H}^2B_{\rm H}$, 
the total power of the BZ jet is \citep[e.g., ][]{BZ77, TNM11}
\begin{eqnarray}\label{L_BZ}
L_{\rm BZ}&\sim&\frac{\kappa}{4\pi c}\Omega_{\rm H}^2\Psi_{\rm BH}^2,
\end{eqnarray}
where $\kappa\approx0.05$, $J$ is the angular momentum of the BH, 
$B_{\rm H}$ is the strength of the magnetic field at the BH, $r_{\rm H}$ is the radius of the BH horizon, 
$c$ is the light speed, and $G$ is the gravitational constant, 
We use $a/M_{\rm BH}\sim0.7$ as a fiducial value \citep{ST06}. 
Taking into account the beaming correction \citep[$\theta_{\rm j}^2\sim10^{-3}$; ][]{Fong+14, Fong+15} 
and the radiative efficiency \citep[$\eta\sim0.1$; ][]{Zha+07}, 
the observed isotropic luminosity is
\begin{eqnarray}\label{L_iso}
L_{\rm iso}\sim\eta(2/\theta_{\rm j}^2)L_{\rm BZ}\sim10^2L_{\rm BZ}.
\end{eqnarray}
As long as the pressure of fallback matter supports the magnetic flux on the BH, 
the BZ power remains flat, $L_{\rm iso}\propto t^0$ \citep[see also][]{TG15}. 

The timescale of each component $T_{\rm i}$ is determined by the pressure balance between  
the fallback matter and magnetic field. 
Such a disk state is the so-called magnetically arrested disk \citep[e.g., ][]{NIA03}. 
If the matter pressure cannot support the magnetic flux on BH, the BZ power reduces. 
The temporal evolution of the mass accretion rate is \citep[e.g., ][]{R07, Kyu+15}, 
\begin{eqnarray}\label{dotM}
\dot{M}=\frac{2}{3}\frac{M_{\rm f}}{0.1~{\rm s}} \left(\frac{t}{0.1{\rm s}}\right)^{-5/3},~~~(t>0.1~{\rm s}),
\end{eqnarray}
where $M_{\rm f}\equiv\int_{0.1~{\rm s}}^{\infty}\dot{M}dt$ is the total fallback mass after the reference time $t>0.1$ s.  
From the force balance between the matter and magnetic field pressures, 
the characteristic timescale of the BZ jet is given by
\begin{eqnarray}\label{sec3:T}
T&\sim&1\times10^4~\left(\frac{M_{\rm f}}{10^{-3}M_{\odot}}\right)^{3/5}\left(\frac{B_{\rm H}}{10^{12}{\rm G}}\right)^{-6/5}~{\rm s}, 
\end{eqnarray}
where we use the BH mass $M_{\rm BH}=3M_{\odot}$, and the radial velocity $v_{\rm R}\sim10^{-2}\sqrt{GM_{\rm BH}/R_{\rm H}}$ 
\citep[e.g., ][]{TNM11, ZCST14}. 

We discuss the implications for the BH engine model \citep{KI15} from the results in Section 3.
Since the light curve implied by the BH engine model is consistent with the observations, 
the results in \citet{KI15} are also applicable to our sample.
Using the obtained parameters $L_{\rm iso, EX}$, $L_{\rm iso, PL}$, $T_{\rm EX}$, and $T_{\rm PL}$, 
we can estimate the magnetic field $B_{\rm H}$ and the total fallback mass $M_{\rm f}$ as follows.
The strength of the magnetic field $B_{\rm H}$ is determined by the observed luminosity 
$L_{\rm iso}$ from Equation (\ref{L_iso}), as
\begin{eqnarray}\label{B_H}
B_{\rm H}&\sim&3\times10^{12}~\left(\frac{\eta/\theta_{\rm j}^2}{10^2}\right)^{-1/2} 
\left(\frac{L_{\rm iso}}{10^{47}{\rm erg~s}^{-1}}\right)^{1/2}{\rm G}.
\end{eqnarray}
On the other hand, the total fallback mass $M_{\rm f}$ is derived from Equations (\ref{sec3:T}) and (\ref{B_H}), as
\begin{eqnarray}\label{M_f}
M_{\rm f}&\sim&1\times10^{-2}\left(\frac{\eta/\theta_{\rm j}^2}{10^2}\right)^{-1} \nonumber \\
& &\times\left(\frac{L_{\rm iso}}{10^{47}{\rm erg~s}^{-1}}\right)\left(\frac{T}{10^4{\rm s}}\right)^{5/3}M_{\odot}.
\end{eqnarray}

Figure \ref{BHmodel} shows the distributions of $B_{\rm H}$ and $M_{\rm f}$ from the observations. 
The ranges of the magnetic field strength are $B_{\rm H}\sim10^{12}-10^{14}$ G for the extended emission, 
and $B_{\rm H}\sim10^{11}-10^{13}$ G for the plateau emission.
The range of the magnetic field for the plateau emission is consistent with that of typical NSs 
and PSR J0737-3039B in the double pulsar system \citep{Lyne+04}.
The range of the fallback mass $M_{\rm f}\sim10^{-5}-10^{-1}M_{\odot}$ is consistent with 
the numerical relativity simulations \citep[e.g., ][]{Hot+13, Kyu+15, Kaw+15, Foucart+15, Foucart+17}. 
There is significant dispersion in the distribution of the derived fallback mass. 
For BH-NS mergers, the total mass and mass ratio of the binaries before the merger 
should have some dispersions, which may explain the dispersion
of the fallback mass $M_{\rm f}$. 
On the other hand, for the NS-NS mergers, the dispersions of the total mass and the mass ratio are relatively narrow. 
Then, the large dispersion could stem from the varieties of the radiation efficiency 
$\eta$ and the jet opening angle $\theta_{\rm j}$. 

Note that the populations of the extended and plateau emission do not overlap in $M_{\rm f}-B_{\rm H}$ plot (Figure \ref{BHmodel}).
This is because the distributions of two durations $T_{\rm EX}$ and $T_{\rm PL}$ are completely separated (Figure \ref{fig:L-T}).
In fact, two populations are split by the line $M_{\rm f}\propto B_{\rm H}^2$ derived from Equation (\ref{sec3:T}) 
with $T=$ const.
From Equation (\ref{sec3:T}), 
the timescale does not depend on the radiation efficiency $\eta$ and the jet opening angle $\theta_{\rm j}$. 
Then, the populations of the extended and plateau emission do not overlap in Figure \ref{BHmodel} 
irrespective of $\eta$ and $\theta_{\rm j}$. 

In some bursts, the fallback mass calculated from the extended emission is 
rather different from that from the plateau emission (Figure \ref{BHmodel}). 
The duration of extended emission $T_{\rm EX}$ could become short 
because of the decrease of the magnetic flux $\Psi_{\rm BH}$ via the magnetic reconnection 
\citep[phase VI in Figure 2 of ][]{KI15}. 
Then, the fallback mass $M_{\rm f}$ from Equation (\ref{M_f}) is underestimated. 
On the other hand, at the plateau emission phase, 
the initial magnetic flux of an NS before the merger would give 
the lower bound on $\Psi_{\rm BH}$ \citep[phase VII in Figure 2 of ][]{KI15}. 
Then, the fallback mass estimated from Equation (\ref{M_f}) tends to be larger than that for the extended emission. 
Therefore, it is reasonable that the fallback mass derived from the extended emission tends to 
be smaller than that from the plateau emission. 

In a few bursts, the temporal flux decay at the end of the extended emission seems much steeper than the BH engine model 
\citep[e.g., ][]{Barthelmy+05}. 
Such steep decays are produced only by the activity of the central engine \citep{IKZ05}.
The light curve model described in Equation (\ref{lightcurve}) 
and shown in Figure \ref{fig:lightcurve} is a simple toy model. 
Detailed processes such as the magnetic flux decay due to the magnetic reconnection near the BH horizon
are not included in our model, and 
can produce the short-timescale seen at the end of the extended emission phase. 
In fact, the magnetic reconnection and the resultant decline of the magnetic flux occurs near the BH horizon, 
so that the minimum decay timescale of the extended emission is 
the light crossing time of the BH horizon ($\ll 1$ s). 

For the light curve, we assume the temporal decaying index $\alpha=40/9$ in Equation (\ref{lightcurve}). 
This value is calculated from the temporal evolution of the mass accretion rate, $\dot{M}\propto t^{-5/3}$ 
\citep[e.g., ][]{R07, Kyu+15}. 
The mass accretion rate from the remnant disk of NS-NS and BH-NS mergers at late phase ($t>1$ s) 
has been studied \citep[e.g., ][]{MPQ08, MAQM10, FM13, MF14, Fer+15, Fer+16}.
If the effect of the disk outflow is negligible as in the low viscosity case, 
the mass accretion rate of the advection-dominated disk scales with $\dot{M}\propto t^{-4/3}$ \citep{MPQ08}. 
Then, the total fallback mass $M_{\rm f}$ required for the duration of the plateau emission is 
about an order of magnitude smaller than the results in Figure \ref{BHmodel}. 
On the other hand, 
the disk winds are powerful and make the time dependence of the accretion rate steepen, 
$\dot{M}\propto t^{-2.2}$ after $t\gtrsim1$ s in some simulation results \citep{Fer+15, Fer+16}.
Then, the required mass $M_{\rm f}$ becomes large, 
an order of solar mass $\sim M_{\odot}$ in some fraction of the bursts. 
In addition, the heating by the decay of $r$-process elements could also affect the accretion rate \citep{MAQM10}. 
Since the current numerical simulations only follow up to $\lesssim 10$ s after the merger \citep{Fer+15, Fer+16} 
and do not include some important effects such as the magnetic field \citep[e.g., ][]{Kiuchi+15, SM17}, 
further studies at late phase ($t\gg 10$ s) are required.

\section{CONCLUSIONS AND DISCUSSION}
\label{sec:discussion}

We obtain the statistical properties of the extended and plateau emission components 
by comparing the phenomenological light curves with  
the observational data of 65 short GRBs detected by {\it Swift}/BAT and XRT.  
The phenomenological light curve in Equation (\ref{lightcurve}) is found to be consistent 
with {\it all} the observations in our sample. 
The number of bursts with both the extended and plateau emission components in our sample is 33, 
which is more than three times larger than that in previous works \citep{GOW14, KI15}. 
Furthermore, the remaining bursts in our sample (32 bursts) may also have both emission components, 
which are consistent with the observations (gray dot-dashed curves in Figure \ref{fig:lightcurve}). 
This is the first statistical indication that the extended and plateau emission components are ubiquitous 
in short GRB light curves. 

The extended and plateau emission are clearly distinct populations 
as shown in Figure \ref{fig:L-T}.
The ranges of the luminosity and duration for the extended emission are $L_{\rm iso, EX}\sim10^{47}-10^{50}$ erg s$^{-1}$ 
and $T_{\rm EX}\sim10^2-10^3$ s, respectively. 
The ranges extend to the lower and longer values than those of the extended emission detected 
in the {\it Swift}/BAT band (Figure \ref{fig:histgram}). 
The ranges of the luminosity and duration for the plateau emission, 
$L_{\rm iso, PL}\sim10^{43}-10^{47}$ erg s$^{-1}$ and $T_{\rm PL}\sim10^4-10^5$ s, respectively, 
also extend to the lower and longer values 
than those in the previous works with a small number of sample \citep[$\sim10$; ][]{Row+13, GOW14, KI15}. 
The isotropic energy of the extended emission $E_{\rm iso, EX}$ is comparable to that of the prompt emission $E_{\rm iso, PR}$. 
The ratio $E_{\rm iso, EX}/E_{\rm iso, PR}\sim0.1-10$ is the same as the short GRBs with the extended emission detected 
in the {\it Swift}/BAT band \citep[e.g., ][]{Per+09}. 
On the other hand, the isotropic energy of the plateau emission, $E_{\rm iso, PL}\sim10^{47}-10^{51}$ erg, 
is $\sim1-100$ times lower than that of the prompt and the extended emission, $E_{\rm iso, PR}$ and $E_{\rm iso, EX}$. 

We consider that the extended emission detected by {\it Swift}/BAT \citep[e.g., ][]{NGS11} is the same component 
detected by {\it Swift}/XRT with duration $\sim10^2$ s.
Actually, the luminosity and duration are continuously distributed (Figure \ref{fig:histgram}).
The fraction of short GRBs with the extended emission in our sample is $\sim$75\%, 
which is much higher than the previous values, 
$\sim2-25$\% in the {\it Swift}/BAT data \citep{NGS10, Sakamoto+11, KBGL15, Lien+16}, 
37.5\% in bright short GRBs detected by both {\it Swift}/BAT and XRT \citep{Kag+15}, 
$\sim5$\% in the {\it Fermi}/GBM data \citep{KBGL15}, and 
$\sim$7\% in the BATSE data \citep{BKG13}, 
although these statistical values should not be directly compared 
because selection criteria are different.  
The energy spectrum of the extended emission is soft compared with 
that of the prompt emission \citep[e.g., ][]{Kag+15, Lien+16}. 
In addition, at early phase of the light curve $\sim10^2$ s, 
most of our samples are detected by {\it Swift}/XRT whose flux sensitivity is much higher than that of {\it Swift}/BAT. 
In fact, the {\it Swift}/BAT-detected extended emission has the higher luminosity and shorter duration
in the entire extended emission population as shown in Figure \ref{fig:histgram}.
These effects would increase the fraction of the XRT-detected bursts with the extended emission.
On the other hand, the starting time of the observation by {\it Swift}/XRT is usually
$60-100$ s after the prompt emission, which is comparable to the duration of the extended emission. 
Then, {\it Swift}/XRT would miss some of the extended emission with short duration, 
and the actual fraction of the short GRBs with the extended emission could be higher than $\sim75$\%. 

The fraction of the short GRBs with plateau emission in our samples is also $\sim$75\%.
Since the plateau emission is generally $\sim10^{-3}-10^{-4}$ times dimmer 
than the extended emission (Figure \ref{fig:L-L_T-T} A), 
the flux of the plateau emission in some bursts would be below the detection limit. 
Actually, the upper limit on the luminosity of the plateau emission in our sample
is typically $\sim10^{44}-10^{45}$ erg s$^{-1}$ (gray dot-dashed curves in Figure \ref{fig:lightcurve}), 
higher than the lower end of the luminosity distribution of plateau emission (Figure \ref{fig:histgram})
at the time $\sim10^5$ s.
Therefore, the actual fraction would be higher than $\sim75$\%, 
and it is consistent with that almost all short GRBs have the plateau emission component.

There is a possible indication for the short- and long-lived populations of the light curves at the plateau phase, 
$t>10^4$ s \citep{SG09}.
From Figure \ref{fig:histgram}(B), the distribution may be composed by two distinct populations, 
$T_{\rm PL}\sim10^4$ s and $T_{\rm PL}\sim10^5$ s, although it is not so obvious in our sample.

The X-ray excess component with timescale $\sim10^6$ s, longer than the plateau emission, 
has also detected by {\it XMM-Newton} and {\it Chandra} in some bursts \citep[e.g., ][]{Fong+14, Fong+16}. 
For the excess component, the activity near the central engine such as an accretion disk is considered 
\citep[e.g., ][]{R07, RB09, Kyu+15, KIN16}. 
For the excess component detected by {\it XMM-Newton} in GRB 130603B, 
the fallback mass $\sim0.02M_{\odot}$ are required to produce 
the observed luminosity $\sim10^{41}$ erg s$^{-1}$ at $\sim7$ day in the rest-frame 
if the radiation efficiency of the accretion disk is $\sim0.1$ and 
the mass accretion rate follows $\dot{M}\propto t^{-5/3}$ at $t>0.1$ s \citep{KIN16}. 
This is consistent with the mass $M_{\rm f}$ from the observed plateau emission parameters 
if the radiation efficiency of the jet $\eta\sim2.5\times10^{-2}$ at the plateau emission phase is assumed. 
From Figure \ref{BHmodel}, a fraction $\sim17\%$ of bursts with the plateau emission requires the fallback mass 
$M_{\rm f}\gtrsim0.02M_{\odot}$, so that these bursts could have the X-ray excess with $L\gtrsim10^{41}$ erg s$^{-1}$ 
at $\sim7$ day in the rest-frame. 
The excess component could also contribute to the energy source of the observed macronovae \citep{KIN16, Jin+16}. 

We use the light curve data from UK {\it Swift} Science Data Center \citep{Eva+07, Eva+09}. 
In some data, the exposure time is much shorter than the error duration. 
We show the data points with the low fractional exposure ($<0.1$) as gray points in Figure \ref{fig:lightcurve}. 
Even if we do not use such data, the results do not significantly change as shown in Figure \ref{fig:lightcurve}. 
For the last data point in some GRBs, even if the fractional exposure is close to unity, the arrival time of
almost all the photons in the bin is clustering in the first short time range compared with the error duration. 
Then, we may overestimate the duration in the light curve fitting. 
A example is the last data point in GRB 090515, which is divided by the detection point with duration 
$<10^2$ s and the upper limit with $\sim10^3$ s in the analysis by \citet{Rowlinson+10a}. 
Note that for GRB 090515, the last data point corresponds to the decay phase of the extended emission (Figure \ref{fig:lightcurve}), 
which could decay more rapidly than the model light curve. 
We consider that such data points do not significantly change our results.

The emission solid angles of the extended and plateau emission are unknown.
If the emission is isotropic, the extended and plateau emission could be easily detectable
as EM counterparts to the NS binary mergers. 
For the {\it Swift}/BAT-detected extended emission, \citet{Nakamura+14} estimated the emission solid angle, 
$\sim10^{-3}$ steradian, by comparing the detection rate by {\it Swift}/BAT with the estimated 
merger rate $\sim10^3$ Gpc$^{-3}$ yr$^{-1}$. 

From our results, the extended emission is mainly detected in the {\it Swift}/XRT band, 
and the luminosity range is $10^{47}$ erg s$^{-1}$ $\lesssim L_{\rm iso, EE}\lesssim 10^{50}$ erg s$^{-1}$. 
The all-sky survey in soft X-ray band is planned by {\it eROSITA}. 
Let us consider the detection rate of the extended emission by {\it eROSITA} as a function of the emission solid angle, 
$\Delta\Omega_{\rm EX}$. 
The flux sensitivity limit of the {\it eROSITA} single survey pass is $\sim10^{-13}$ erg cm$^{-2}$ s$^{-1}$ \citep{Mer+12}, 
which corresponds to the detection horizon $\sim100$ Gpc in the luminosity distance and the comoving volume $4\times10^3$ Gpc$^3$ 
for the luminosity $\sim10^{47}$ erg s$^{-1}$.  
Then, the event rate in the volume is $\sim0.1$ s$^{-1}$ for the NS binary merger rate $R_{\rm merger}\sim10^3$ Gpc$^{-3}$ yr$^{-1}$. 
Using the typical duration $\sim100$ s, there are always $\sim10$ events in the all-sky. 
Since a field-of-view of {\it eROSITA} is $0.833$ deg$^2$ and the scanning speed is 
one full circle per four hours \citep{Mer+12}, 
the detection rate is relatively high $\sim0.5(\Delta\Omega_{\rm EX}/4\pi)(R_{\rm merger}/10^3~{\rm Gpc}^{-3}~{\rm yr}^{-1})$ day$^{-1}$. 
The events will be detected only in one scanning, in contrast to stationary sources.
Therefore, {\it eROSITA} could significantly constrain the emission solid angle of the extended emission. 

If the extended emission is isotropic, off-axis events should 
appear as short GRB-less X-ray flashes or long GRBs with relatively simple shaped light 
curves and unusual host galaxy properties compared to normal long GRBs. XRF GRB060428b, 
which had a light curve similar to an observed X-ray flash and which was localized to 
a potential host elliptical galaxy \citep{Perley+07}, 
was suggested as a possible off-axis extended emission of short GRB \citep{MQT08}. 
We will investigate long GRB afterglows with similar light curves to short GRBs in future work.

In the light curves of long GRBs, the existence of a single long-lasting component, so-called the plateau emission 
in addition to the prompt emission and X-ray flares, has been established 
\citep[e.g., ][]{OBrien+06, Willingale+07, GNGC09, Y09, Grupe+13, Rowlinson+14, Dainotti+15, Wang+15}.
The duration distributions of the extended and plateau emission in short GRBs 
(Figure \ref{fig:L-L_T-T} B) are similar to those of the prompt and plateau emission in long GRBs, 
respectively \citep[e.g., ][]{Willingale+07}. 
In addition, the distribution of the fluence ratio between the extended and plateau emission 
in short GRBs (Figure \ref{fig:EE-PL} A) is also similar to that of the ratio 
between the prompt and plateau emission in long GRBs \citep[e.g., ][]{Willingale+07}. 
These similarities may suggest that the physical conditions of the model of \citet{KI15} are realized 
in both short and long GRBs. 
In fact, the central engine activities, mass ejection, and fallback accompanied by supernova explosion 
are also expected in long GRBs.

There are some caveats to be addressed in the future work. 
First, we fit the phenomenological light curve with the observations by eye inspection, 
because the light curves often have additional complex structures of X-ray flares \citep[e.g., ][]{Margutti+11}.  
The observed data of most bursts would be insufficient to separate such complex structures. 
In order to separate the additional structures from the flat component, 
a sample with more sufficient X-ray data is required. 

Second, we also neglect the contribution of the normal afterglow in the X-ray light curve observed by {\it Swift}. 
Although the normal afterglow may contribute to some bursts whose light curves are consistent with a single power-law form 
\citep[e.g., ][]{Lu+15}, 
these light curves are also able to fit with the extended and plateau emissions as demonstrated in this paper. 
The isotropic energies of the extended and plateau emission components are almost comparable to that of the prompt emission, 
$E_{\rm iso, EX}/E_{\rm iso, PR}\sim0.1-10$ and $E_{\rm iso, PL}/E_{\rm iso, PR}\sim0.01-1$, as shown in Figure \ref{fig:EE-PL}, 
which would also suggest the continuous energy injection from the central engine. 
Note that most of the {\it Swift}/XRT data are within $\lesssim10^5$ s after the prompt emission, 
so that the normal afterglow would significantly contribute to the X-ray light curve 
at later time ($\gtrsim10^5$ s), at which the jet break is seen \citep[e.g., ][]{Fong+15}.  

Third, we use the specific value $\alpha=40/9$ for the phenomenological light curve. 
Although the parameters $L_{\rm iso, EX}, L_{\rm iso, PL}, T_{\rm EX}$, and $T_{\rm PL}$ do not significantly depend on $\alpha$
(see the dotted lines ($\alpha=5/3$) in Figure \ref{fig:lightcurve}), 
the light curve with small $\alpha$ could explain the observed data using a single emission component 
in a few bursts. 
Then, the fraction of bursts with the plateau emission would slightly decrease. 
On the other hand, the light curve model with small $\alpha$ is not consistent 
with some bursts with the rapidly decaying plateau emission. 
In the theoretical point of view, 
the parameter $\alpha$ depends on the accretion model as discussed in Section \ref{sec:BHmodel}. 
A sample with sufficient X-ray data will provide the fitted value of $\alpha$ to characterize the light curve 
and a clue to the fallback to the central engine. 

Recently, a new X-ray transient detected by {\it Chandra} has been reported \citep{Bauer+17}.
Using the measured redshift $z\sim2.2$ \citep{Bauer+17},  
the luminosity $\sim10^{47}$ erg s$^{-1}$ and duration $\sim10^2$ s could be consistent with 
the extended emission properties. 
The decaying light curve of the X-ray transient seems to become gradually shallow (in their Figure 4), 
so that the plateau emission may also contribute to the observed light curves.
For the prompt emission properties, the interplanetary network \citep[IPN; ][]{Atteia+87} gives the limits on the fluence and
peak photon flux \citep[$<10^{-6}$ erg cm$^{-1}$ and $<1$ photon cm$^{-2}$ s$^{-1}$, respectively, 
in the 25-150 keV range ;][]{Bauer+17}.
A fraction of the prompt emission of short GRBs is fainter than the limits \citep{Lien+16}. 
Within the 2$\sigma$ range of the measured photometric redshift, 
the host galaxy could locate at $z\sim0.39$ \citep{Bauer+17}. 
If we use this value, the luminosity $\sim10^{44}$ erg s$^{-1}$ and duration $\sim10^{3}$ s are consistent 
with those of the scattered plateau emission. 
In addition, the event rate of the transient is $\sim10^2-10^3$ yr$^{-1}$ Gpc$^{-3}$ in the low $z(\lesssim1)$ case, 
which is also consistent with the rate of the orphan short GRBs \citep{B14, Bauer+17}. 

Possible macronova emission was reported in short GRB 160821B \citep{Tanvir+17, Troja+16b, Kasliwal+17}, 
which occurred at $z\sim0.16$ \citep{Levan+16}, closer than GRB 130603B. 
The peak luminosity is comparable to that of GRB 130603B while the peak time is earlier, $t\sim3$ day. 
The observed luminosity is consistent with the expected bolometric luminosity, 
$\sim10^{41}(t/3~{\rm day})^{-2}$ erg s$^{-1}$, in the case of ejecta heated by the plateau activity \citep{KIN15}, 
where we use $\eta=0.1$, $\theta_{\rm j}=0.1$, and the observed value $L_{\rm iso, PL}T_{\rm PL}^2$ in GRB 160821B, 
which is $\sim30$ times smaller than that of GRB 130603B. 
Note that the possible X-ray excess component ($\lesssim10^{42}$ erg s$^{-1}$ at $10^6$ s by {\it Swift}/XRT, 
see Figure \ref{fig:lightcurve}) could also contribute to heating the ejecta \citep{KIN16}. 
Therefore, the engine-powered macronova scenario is consistent with the observations so far. 

\acknowledgments
We are grateful to the anonymous referee for constructive comments. 
We would like to thank Amy Lien for kindly providing the data of the short pulse fluence of short GRBs 
with extended emission, 
and Yutaka Ohira, Masaomi Tanaka, and Ryo Yamazaki for fruitful discussions. 
This work is supported by KAKENHI 16J06773 (S.K.), 24103006, 26247042, 26287051, 
17H01126, 17H06131, 17H06357 (K.I.), and 17H06362 (K.I., T.S.).


\begin{thebibliography}{99}

\bibitem[Abbott et al. (2016a)]{Abbott+16a}
Abbott, B. P., Abbott, R., Abbott, T. D., et al. 2016a, PhRvL, 116, 061102

\bibitem[Abbott et al. (2016b)]{Abbott+16b}
Abbott, B. P., Abbott, R., Abbott, T. D., et al. 2016b, PhRvL, 116, 241103

\bibitem[Abbott et al. (2016c)]{Abbott+16c}
Abbott, B. P., Abbott, R., Abbott, T. D., et al. 2016c, PhRvX, 6, 041015

\bibitem[Abbott et al. (2016d)]{Abbott+16d}
Abbott, B. P., Abbott, R., Abbott, T. D., et al. 2016d, ApJL, 832, L21

\bibitem[Abbott et al. (2017)]{Abbott+17}
Abbott, B. P., Abbott, R., Abbott, T. D., et al. 2017, PhRvL, 118, 221101

\bibitem[Adriani et al. (2016)]{Adriani+16} 
Adriani, O., Akaike, Y., Asano, K., et al. 2016, ApJL, 829, L20

\bibitem[Atteia et al. (1987)]{Atteia+87}
Atteia, J.-L., Barat, C., Hurley, K., et al. 1987, ApJS, 64, 305

\bibitem[Barkov \& Pozanenko (2011)]{BP11}
Barkov, M. V., \& Pozanenko, A. S. 2011, MNRAS, 417, 2161

\bibitem[Barthelmy et al. (2005)]{Barthelmy+05}
Barthelmy, S. D., Cannizzo, J. K., Gehrels, N., et al. 2005, ApJL, 635, L133

\bibitem[Bauer et al. (2017)]{Bauer+17}
Bauer, F. E., Treister, E., Schawinski, K., et al. 2017, MNRAS, 467, 4841

\bibitem[Berger (2006)]{B06}
Berger, E. 2006, GCN, 5952, 1

\bibitem[Berger (2009)]{B09}
Berger, E. 2009, ApJ, 690, 231

\bibitem[Berger (2014)]{B14}
Berger, E. 2014, ARA\&A, 52, 43

\bibitem[Berger et al. (2013a)]{BFC13}
Berger, E., Fong, W., \& Chornock, R. 2013a, ApJL, 774, L23

\bibitem[Berger et al. (2007)]{Berger+07}
Berger, E., Fox, D. B., Price, P. A., et al. 2007, ApJ, 664, 1000

\bibitem[Berger et al. (2005)]{Berger+05}
Berger, E., Price, P. A., Cenko, S. B., et al. 2005, Natur, 438, 988

\bibitem[Berger et al. (2013b)]{Berger+13}
Berger, E., Zauderer, B. A., Levan, A., et al. 2013b, ApJ, 765, 121

\bibitem[Blandford \& Znajek (1977)]{BZ77}
Blandford, R. D., \& Znajek, R. L. 1977, MNRAS, 179, 433

\bibitem[Bostanc\i\ et al. (2013)]{BKG13}
Bostanc\i, Z. F., Kaneko, Y., \& G\"o\u{g}\"u\c{s}, E. 2013, MNARS, 428, 1623

\bibitem[Bromberg et al. (2011)]{BNPS11}
Bromberg, O., Nakar, E., Piran, T., \& Sari, R. 2011, ApJ, 740, 100

\bibitem[Bucciantini et al. (2012)]{BMTQ12}
Bucciantini, N., Metzger, B. D., Thompson, T. A., \& Quataert, E. 2012, MNRAS, 419, 1537

\bibitem[Camp et al. (2013)]{Camp+13}
Camp, J., Barthelmy, S. D., Blackburn, L., et al. 2013, ExA, 36, 505

\bibitem[Castro-Tirado et al. (2015)]{Castro-Tirado+15}
Castro-Tirado, A. J., Sanchez-Ramirez, R., Lombardi, G., et al. 2015, GCN, 17758, 1

\bibitem[Cenko et al. (2008)]{Cenko+08}
Cenko, S. B., Berger, E., Nakar, E., et al. 2008, arXiv:0802.0874

\bibitem[Chen \& Holz (2015)]{CH15}
Chen, H.-Y., \& Holz, D. E. 2015, arXiv:1509.00055

\bibitem[Chornock \& Fong (2015)]{CF15}
Chornock, R., \& Fong, W. 2015, GCN, 17358, 1

\bibitem[Chornock et al. (2013)]{CLB13}
Chornock, R., Lunnan, R., \& Berger, E. 2013, GCN, 15307, 1

\bibitem[Cucchiara \& Levan (2016)]{CL16}
Cucchiara, A., \& Levan, A. J. 2016, GCN, 19565, 1

\bibitem[D'Avanzo et al. (2009)]{D'Avanzo+09}
D'Avanzo, P., Malesani, D., Covino, S., et al. 2009, A\&A, 498, 711

\bibitem[D'Avanzo et al. (2014)]{D'Avanzo+14}
D'Avanzo, P., Salvaterra, R., Bernardini, M. G., et al. 2014, MNRAS, 442, 2342

\bibitem[Dainotti et al. (2015)]{Dainotti+15}
Dainotti, M., Petrosian, V., Willngale, R., et al. 2015, MNRAS, 451, 3898

\bibitem[de Ugarte Postigo et al. (2014)]{deUgartePostigo+14}
de Ugarte Postigo, A., Th\"one, C. C., Rowlinson, A., et al. 2014, A\&A, 563, A62

\bibitem[Della Valle et al. (2006)]{DellaValle+06}
Della Valle, M., Chincarini, G., Panagia, N., et al. 2006, Natur, 444, 1050

\bibitem[Evans et al. (2007)]{Eva+07}
Evans, P. A., Beardmore, A. P., Page, K. L., et al. 2007, A\&A, 469, 379

\bibitem[Evans et al. (2009)]{Eva+09}
Evans, P. A., Beardmore, A. P., Page, K. L., et al. 2009, MNRAS, 397, 1177

\bibitem[Fan et al. (2017)]{FMH17}
Fan, X., Messenger, C., \& Heng, I. S. 2017, arXiv:1706.05639

\bibitem[Fern\'andez \& Metzger (2013)]{FM13}
Fern\'andez, R., \& Metzger, B. D. 2013, MNRAS, 435, 502

\bibitem[Fern\'andez et al. (2015)]{Fer+15}
Fern\'andez, R., Quataert, E., Schwab, J., Kasen, D., \& Rosswog, S. 2015, MNRAS, 449, 390

\bibitem[Fern\'andez et al. (2017)]{Fer+16}
Fern\'andez, R., Foucart, F., Kasen, D., et al. 2017, CQG, 34, 154001

\bibitem[Fong et al. (2013)]{Fong+13}
Fong, W., Berger, E., Chornock, R., et al. 2013, ApJ, 769, 56

\bibitem[Fong et al. (2011)]{Fong+11}
Fong, W., Berger, E., Chornock, R., et al. 2011, ApJ, 730, 26

\bibitem[Fong et al. (2015)]{Fong+15}
Fong, W., Berger, E., Margutti, R., \& Zauderer, B. A. 2015, ApJ, 815, 102

\bibitem[Fong et al. (2014)]{Fong+14}
Fong, W., Berger, E., Metzger, B. D., et al. 2014, ApJ, 780, 118

\bibitem[Fong et al. (2016)]{Fong+16}
Fong, W., Margutti, R., Chornock, R., et al. 2016, ApJ, 833, 151

\bibitem[Foucart et al. (2017)]{Foucart+17}
Foucart, F., Desai, D., Brege, W., et al. 2017, CQG, 34, 044002

\bibitem[Foucart et al. (2015)]{Foucart+15}
Foucart, F., O'Connor, E., Roberts, L., et al. 2015, PhRvD, 91, 124021

\bibitem[Fraley \& Raftery (2002)]{FR02}
Fraley, C., \& Raftery, A. E. 2002, JASA, 97, 611

\bibitem[Gao \& Fan (2006)]{GF06}
Gao, W.-H., \& Fan, Y.-Z. 2006, ChJAA, 6, 513

\bibitem[Gehrels et al. (2006)]{Gehrels+06}
Gehrels, N., Norris, J. P., Barthelmy, S. D., et al. 2006, Natur, 444, 1044

\bibitem[Ghisellini et al. (2009)]{GNGC09}
Ghisellini, G., Nardini, M., Ghirlanda, G., \& Celotti, A. 2009, MNRAS, 393, 253

\bibitem[Gibson et al. (2017)]{GWGO17}
Gibson, S. L., Wynn, G. A., Gompertz, B. P., \& O'Brien, P. T. 2017, arXiv:1706.04802

\bibitem[Gompertz et al. (2014)]{GOW14}
Gompertz, B. P., O'Brien, P. T., \& Wynn, G. A. 2014, MNRAS, 438, 240

\bibitem[Gompertz et al. (2013)]{GOWR13}
Gompertz, B. P., O'Brien, P. T., Wynn, G. A., \& Rowlinson, A. 2013, MNRAS, 431, 1745

\bibitem[Gottlieb et al. (2017)]{GNP17}
Gottlieb, O., Nakar, E., \& Piran, T. 2017, arXiv:1705.10797

\bibitem[Grupe et al. (2013)]{Grupe+13}
Grupe, D., Nousek, J. A., Veres, P., Zhang, B.-B., \& Gehrels, N. 2013, ApJS, 209, 20

\bibitem[Hartigan (1985)]{H85}
Hartigan, P.~M., 1985, Appl. Stat., 34, 320

\bibitem[Hotokezaka et al. (2013)]{Hot+13}
Hotokezaka, K., Kiuchi, K., Kyutoku, K., et al. 2013, PhRvD, 87, 024001

\bibitem[Hotokezaka \& Piran (2015)]{HP15}
Hotokezaka, K., \& Piran, T. 2015, MNRAS, 450, 1430

\bibitem[Ioka et al. (2005)]{IKZ05}
Ioka, K., Kobayashi, S., \& Zhang, B. 2005, ApJ, 631, 429

\bibitem[Jin et al. (2016)]{Jin+16}
Jin, Z.-P., Hotokezaka, K., Li, X., et al. 2016, NatCo, 7, 12898

\bibitem[Kagawa et al. (2015)]{Kag+15}
Kagawa, Y., Yonetoku, D., Sawano, T., Toyanago, A., Nakamura, T., Takahashi, K., Kashiyama, K., \& Ioka, K. 2015, ApJ, 811, 4

\bibitem[Kaneko et al. (2015)]{KBGL15}
Kaneko, Y., Bostanc\i, Z. F., G\"o\u{g}\"u\c{s}, E., \& Lin, L. 2015, MNARS, 452, 824

\bibitem[Kanner et al. (2012)]{Kanner+12}
Kanner, J., Camp, J., Racusin, J., Gehrels, N., \& White, D. 2012, ApJ, 759, 22

\bibitem[Kasen et al. (2013)]{KBB13}
Kasen, D., Badnell, N. R., \& Barnes, J. 2013, ApJ, 774, 25

\bibitem[Kasliwal et al. (2017)]{Kasliwal+17}
Kasliwal, M. M., Korobkin, O., Lau, R. M., Wollaeger, R., \& Fryer, C. L. 2017, ApJL, 843, L34

\bibitem[Kawaguchi et al. (2015)]{Kaw+15}
Kawaguchi, K., Kyutoku, K., Nakano, H., Okawa, H., Shibata, M., \& Taniguchi, K. 2015, PhRvD, 92, 024014

\bibitem[Kisaka \& Ioka (2015)]{KI15}
Kisaka, S., \& Ioka, K. 2015, ApJL, 804, L16

\bibitem[Kisaka et al. (2015b)]{KIN15}
Kisaka, S., Ioka, K., \& Nakamura, T. 2015b, ApJL, 809, L8

\bibitem[Kisaka et al. (2016)]{KIN16}
Kisaka, S., Ioka, K., \& Nakar, E. 2016, ApJ, 818, 104

\bibitem[Kisaka et al. (2015a)]{KIT15}
Kisaka, S., Ioka, K., \& Takami, H. 2015a, ApJ, 802, 119

\bibitem[Kiuchi et al. (2015)]{Kiuchi+15}
Kiuchi, K., Sekiguchi, Y., Kyutoku, K., Shibata, M., Taniguchi, K., \& Wada, T. 2015, PhRvD, 92, 064034

\bibitem[Kouveliotou et al. (1993)]{Kou+93}
Kouveliotou, C., Meegan, C. A., Fishman, G. J., et al. 1993, ApJL, 413, L101

\bibitem[Krimm et al. (2013)]{Krimm+13}
Krimm, H. A., Holland, S. T., Corbet, R. H. D., et al. 2013, ApJS, 209, 14

\bibitem[Kulkarni (2005)]{K05}
Kulkarni, S. R. 2005, astro-ph/0510256

\bibitem[Kyutoku et al. (2015)]{Kyu+15}
Kyutoku, K., Ioka, K., Okawa,~H., Shibata,~M., \& Taniguchi,~K. 2015, PhRvD, 92, 044028

\bibitem[Lazzati et al. (2016)]{LDMW16}
Lazzati, D., Deich, A., Morsony, B. J., \& Workman, J. C. 2016, arXiv:1610.01157

\bibitem[Lee \& Ramirez-Ruiz (2007)]{LR07}
Lee, W. H., \& Ramirez-Ruiz, E. 2007, NJPh, 9, 17

\bibitem[Leibler \& Berger (2010)]{LB10}
Leibler, C. N., \& Berger, E. 2010, ApJ, 725, 1202

\bibitem[Levan et al. (2016)]{Levan+16}
Levan, A. J., Wiersema, K., Tanvir, N. R., et al. 2016, GCN, 19846, 1

\bibitem[Levesque et al. (2010)]{Levesque+10}
Levesque, E. M., Bloom, J. S., Butler, N. R., et al. 2010, MNRAS, 401, 963

\bibitem[Li \& Paczy\'nsky (1998)]{LP98}
Li, L.-X., \& Paczy\'nsky, B. 1998, ApJL, 507, L59

\bibitem[Lien et al. (2016)]{Lien+16}
Lien, A., Sakamoto, T., Barthelmy, S. D., et al. 2016, ApJ, 829, 7

\bibitem[L\"u et al. (2015)]{Lu+15}
L\"u, H.-J., Zhang, B., Lei, W.-H., Li, Y., \& Lasky, P. D. 2015, ApJ, 805, 89

\bibitem[L\"u et al. (2017)]{Lu+17}
L\"u, H.-J., Zhang, H.-M., Zhong, S.-Q., et al. 2017, ApJ, 835, 181

\bibitem[Lyne et al. (2004)]{Lyne+04}
Lyne, A. G., Burgay, M., Kramer, M., et al. 2004, Sci, 303, 1153

\bibitem[Lyutikov (2013)]{L13}
Lyutikov, M. 2013, ApJ, 768, 63

\bibitem[Malesani et al. (2015)]{Malesani+15}
Malesani, D., Kruehler, T., Xu, D., et al. 2015, GCN, 17755, 1

\bibitem[Margutti et al. (2011)]{Margutti+11}
Margutti, R., Chincarini, G., Granot, J., et al. 2011, MNRAS, 417, 2144

\bibitem[McBreen et al. (2010)]{McBreen+10}
McBreen, S., Kr\"uhler, T., Rau, A., et al. 2010, A\&A, 516, A71

\bibitem[Merloni et al. (2012)]{Mer+12}
Merloni, A., Predehl, P., Becker, W., et al. 2012, arXiv:1209.3114

\bibitem[Metzger et al. (2010a)]{MAQM10}
Metzger, B. D., Arcones, A., Quataert, E., \& Mart\'inez-Pinedo, G. 2010a, MNRAS, 402, 2771

\bibitem[Metzger \& Berger (2012)]{MB12}
Metzger, B. D., \& Berger, E. 2012, ApJ, 746, 48

\bibitem[Metzger \& Fern\'andez (2014)]{MF14}
Metzger, B. D., \& Fern\'andez, R. 2014, MNRAS, 441, 3444

\bibitem[Metzger et al. (2010b)]{Metzger+10}
Metzger, B. D., Mart\'inez-Pinedo, G., Darbha, S., Quataert, E., Arcones, A., Kasen, D., Thomas, R., Nugent, P., Panov, I. V., \& Zinner, N. T. 2010b, MNRAS, 406, 2650

\bibitem[Metzger \& Piro (2014)]{MP14}
Metzger, B. D. \& Piro, A. L. 2014, MNRAS, 439, 3916

\bibitem[Metzger et al. (2008a)]{MPQ08}
Metzger, B. D., Piro, A. L., \& Quataert, E. 2008a, MNRAS, 390, 781

\bibitem[Metzger et al. (2008b)]{MQT08}
Metzger, B. D., Quataert, E., \& Thompson, T. A. 2008b, MNRAS, 385, 1455

\bibitem[Nagakura et al. (2014)]{Nagakura+14}
Nagakura, H., Hotokezaka, K., Sekiguchi, Y., Shibata, M., \& Ioka, K. 2014, ApJL, 784, L28

\bibitem[Nakamura et al. (2014)]{Nakamura+14}
Nakamura, T., Kashiyama, K., Nakauchi, D., et al. 2014, ApJ, 796, 13

\bibitem[Nakar (2007)]{N07}
Nakar, E. 2007, PhR, 442, 166

\bibitem[Narayan et al. (2003)]{NIA03}
Narayan, R., Igumenshchev, I. V., \& Abramowicz, M. A. 2003, PASJ, 55, L69

\bibitem[Narayan et al. (1992)]{NPP92}
Narayan, R., Paczy\'nsky, B., \& Piran, T. 1992, ApJL, 395, L83

\bibitem[Narayana Bhat et al. (2016)]{Bhat+16}
Narayana Bhat, P., Meegan, C. A., von Kienlin, A., et al. 2016, ApJS, 223, 28

\bibitem[Norris \& Bonnell (2006)]{NB06}
Norris, J. P., \& Bonnell, J. T. 2006, ApJ, 643, 266

\bibitem[Norris et al. (2011)]{NGS11}
Norris, J. P., Gehrels, N., \& Scargle, J. D. 2011, ApJ, 735, 23

\bibitem[Norris et al. (2010)]{NGS10}
Norris, J. P., Gehrels, N., \& Scargle, J. D. 2010, ApJ, 717, 411

\bibitem[O'Brien et al. (2006)]{OBrien+06}
O'Brien, P. T., Willingale, R., Osborne, J., et al. 2006, ApJ, 647, 1213

\bibitem[Perley et al. (2007)]{Perley+07}
Perley, D. A., Bloom, J. S., Butler, N. R., Li, W., \& Chen, H.-W. 2007, in AIP Conf. Proc. 937, Supernova 1987A: 20 Years After: Supernovae and Gamma-Ray Bursters, ed. S. Immler, K. Weiler, \& R. McCray, 526

\bibitem[Perley et al. (2008)]{Perley+08}
Perley, D. A., Bloom, J. S., Modjaz, M., et al. 2008, GCN, 7889, 1

\bibitem[Perley et al. (2009)]{Per+09}
Perley, D. A., Metzger, B. D., Granot, J., et al. 2009, ApJ, 696, 1871

\bibitem[Prochaska et al. (2005)]{Prochaska+05}
Prochaska, J. X., Bloom, J. S., Chen, H.-W., et al. 2005, GCN, 3399, 1

\bibitem[Rossi \& Begelman (2009)]{RB09}
Rossi, E. M., \& Begelman, M. C. 2009, MNRAS, 392, 1451

\bibitem[Rosswog (2007)]{R07}
Rosswog, S. 2007, MNRAS, 376, L48

\bibitem[Rowlinson et al. (2014)]{Rowlinson+14}
Rowlinson, A., Gompertz, B. P., Dainotti, M., et al. 2014, MNRAS, 443, 1779

\bibitem[Rowlinson et al. (2013)]{Row+13}
Rowlinson, A., O'Brien, P. T., Metzger, B. D., Tanvir, N. R., \& Levan, A. J. 2013, MNRAS, 430, 1061

\bibitem[Rowlinson et al. (2010a)]{Rowlinson+10a}
Rowlinson, A., O'Brien, P. T., Tanvir, N. R., et al. 2010a, MNRAS, 409, 531

\bibitem[Rowlinson et al. (2010b)]{Rowlinson+10}
Rowlinson, A., Wiersema, K., Levan, A. J., et al. 2010b, MNRAS, 408, 383

\bibitem[Sakamoto et al. (2011)]{Sakamoto+11}
Sakamoto, T., Barthelmy, S. D., Baumgartner, W. H., et al. 2011, ApJS, 195, 27

\bibitem[Sakamoto \& Gehrels (2009)]{SG09}
Sakamoto, T., \& Gehrels, N. 2009, in AIP Conf. Proc. 1133, Gamma-Ray Bursts, 6th Huntsville Symposium, ed. C. Meegan, N. Gehrels, \& C. Kouveliotou (Melville, NY: AIP), 112

\bibitem[Sakamoto et al. (2013)]{Sakamoto+13}
Sakamoto, T., Troja, E., Aoki, K., et al. 2013, ApJ, 766, 41

\bibitem[Sari et al. (1998)]{SPN98}
Sari, R., Piran, T., \& Narayan, R. 1998, ApJL, 497, L17

\bibitem[Savchenko et al. (2016)]{Savchenko+16}
Savchenko, V., Ferrigno, C., Mereghetti, S., et al. 2016, ApJL, 820, L36

\bibitem[Shibata \& Taniguchi (2006)]{ST06}
Shibata, M., \& Taniguchi, K. 2006, PhRvD, 73, 064027

\bibitem[Siegel \& Metzger (2017)]{SM17}
Siegel, D. M., \& Metzger, B. D. 2017, arXiv:1705.05473

\bibitem[Singer \& Price (2016)]{SP16}
Singer, L. P., \& Price, L. R. 2016, PhRvD, 93, 024013

\bibitem[Singer et al. (2014)]{Singer+14}
Singer, L. P., Price, L. R., Farr, B., et al. 2014, ApJ, 795, 105

\bibitem[Soderberg et al. (2006)]{Soderberg+06}
Soderberg, A. M., Berger, E., Kasliwal, M., et al. 2006, ApJ, 650, 261

\bibitem[Sugizaki et al. (2011)]{Sugizaki+11}
Sugizaki, M., Mihara, T., Serino, M., et al. 2011, PASJ, 63, 635

\bibitem[Sun et al. (2017)]{SZG16}
Sun, H., Zhang, B., \& Gao, H. 2017, ApJ, 835, 7

\bibitem[Tanaka \& Hotokezaka (2013)]{TH13}
Tanaka, M., \& Hotokezaka, K. 2013, ApJ, 775, 113

\bibitem[Tanvir et al. (2017)]{Tanvir+17}
Tanvir, N. R., et al. 2017, The Physics of Extreme-Gravity Stars, Stockholm, Sweden

\bibitem[Tanvir et al. (2013)]{Tanvir+13}
Tanvir, N. R., Levan, A. J., Fruchter, A. S., et al. 2013, Natur, 500, 547

\bibitem[Tchekhovskoy \& Giannios (2015)]{TG15}
Tchekhovskoy, A., \& Giannios, D. 2015, MNRAS, 447, 327

\bibitem[Tchekhovskoy et al. (2011)]{TNM11}
Tchekhovskoy, A., Narayan, R., \& McKinney, J. C. 2011, MNRAS, 418, L79

\bibitem[Thoene et al. (2010)]{Thoene+10}
Thoene, C. C., de Ugarte Postigo, A., Vreeswijk, P. 2010, GCN, 10971, 1

\bibitem[Troja et al. (2016a)]{Troja+16}
Troja, E., Sakamoto, T., Cenko, S. B., et al. 2016a, ApJ, 827, 102

\bibitem[Troja et al. (2016b)]{Troja+16b}
Troja, E., Tanvir, N., Cenko, S. B., et al. 2016b, GCN, 20222, 1

\bibitem[Usov (1992)]{U92}
Usov, V. V. 1992, Natur, 357, 472

\bibitem[Wang et al. (2015)]{Wang+15}
Wang, X.-G., Zhang, B., Liang, E.-W., et al. 2015, ApJS, 219, 9

\bibitem[Willingale et al. (2007)]{Willingale+07}
Willingale, R., O'Brien, P. T., Osborne, J. P., 2007, ApJ, 662, 1093

\bibitem[Yamazaki (2009)]{Y09}
Yamazaki, R. 2009, ApJL, 690, L118

\bibitem[Yu, Zhang \& Gao (2013)]{YZG13}
Yu, Y.-W., Zhang, B., \& Gao, H. 2013, ApJL, 776, L40

\bibitem[Yuan et al. (2015)]{Yuan+15}
Yuan, W., Zhang, C., Feng, H., et al. 2015, arXiv:1506.07735

\bibitem[Zamaninasab et al. (2014)]{ZCST14}
Zamaninasab, M., Clausen-Brown, E., Savolainen, T., \& Tchekhovskoy, A. 2014, Natur, 510, 126

\bibitem[Zhang et al. (2007)]{Zha+07}
Zhang, B., Liang, E., Page, K. L., et al. 2007, ApJ, 655, 989

\bibitem[Zhang \& M\'esz\'aros (2001)]{ZM01}
Zhang, B., \& M\'esz\'aros, P. 2001, ApJL, 552, L35

\end{thebibliography}
\end{document}